\documentclass[a4paper,11pt]{article}
\pdfoutput=1
\usepackage{jheppub} % see the JHEP-author-manual
\usepackage{amsmath,amssymb,mathtools,bbm}
\usepackage[T1]{fontenc}          % output font encoding
\usepackage{booktabs,tabularx}
\usepackage{graphicx}
\usepackage{xspace}
\usepackage{tikz}
\usetikzlibrary{arrows,calc,matrix,positioning,shapes,trees}
\usetikzlibrary{calc,through}				% For coordinates
\usepackage{tikz-cd}
\usepackage{pgffor}	
\usepackage{listings,enumerate}
\usepackage[absolute]{textpos}
\usepackage[many]{tcolorbox}
\usepackage{xparse}
\usepackage{slashed}
\usepackage{calligra}
\usepackage[font=small,labelfont=bf,format=plain,margin=0.05\textwidth]{caption}
\usepackage{url,hyperref}
\usepackage[utf8]{inputenc}

\usetikzlibrary{decorations.pathmorphing}	% For Feynman Diagrams
\usetikzlibrary{decorations.markings}
\tikzset{
	 >=stealth', %%  Uncomment for more conventional arrows
	fermionloop/.style={draw=black, postaction={decorate},
		decoration={markings,mark=at position .58 with {\arrow[draw=black,rotate=11]{Latex}}}},
	fermion/.style={draw=black, postaction={decorate},
		decoration={markings,mark=at position .58 with {\arrow[draw=black,rotate=0]{Latex}}}},
    fermionnoarrow/.style={draw=black},
    gluon/.style={decorate, draw=black,
        decoration={coil,aspect=1. ,amplitude=3.pt, segment length=6pt}},
    scalar/.style={dashed,draw=black, postaction={decorate},
        decoration={markings,mark=at position .55 with {\arrow[draw=black]{>}}}},
    scalarnoarrow/.style={dashed,draw=black},
    ghost/.style={densely dotted,draw=black, postaction={decorate},
    	decoration={markings,mark=at position .55 with {\arrow[draw=black,rotate=0]{Latex}}}},
}

\tikzstyle{block} = [draw, rectangle, 
minimum height=3em, minimum width=6em]
\tikzset{cross/.style={cross out, draw, 
		minimum size=2*(#1-\pgflinewidth), 
		inner sep=0pt, outer sep=0pt}}

\title{\Large Resummation of terms enhanced by trilinear squark-Higgs couplings in the MSSM}
\author[a]{Thomas Kwasnitza,}
\author[a]{Dominik St\"ockinger}

\affiliation[a]{Institut f\"ur Kern- und Teilchenphysik,
	TU Dresden,\\ Zellescher Weg 19, 01069 Dresden, Germany}

\emailAdd{thomas.kwasnitza@mailbox.tu-dresden.de}
\emailAdd{dominik.stoeckinger@tu-dresden.de}

\abstract{
We analyze the appearance of  the trilinear squark-Higgs couplings $x_q$ in
Green functions in the Higgs sector of the MSSM and in threshold
corrections to the SM. The results are constraints on maximal powers of
$x_q$ in QCD-related loop corrections. In practice these often imply
all-order resummations of leading or subleading $x_q$ contributions in
SM-parametrized expressions.
We present a variety of
all-order resummation relations for $\Delta \lambda$ which include
such $x_q$-enhanced terms and different orders in Yukawa and gauge couplings. We contrast which terms cannot be resummed.
}

\keywords{Higgs, Mass, MSSM}

\arxivnumber{}

\hypersetup{
	pdftitle    = {Resummation of terms enhanced by trilinear squark-Higgs couplings in the MSSM},
	pdfauthor   = {Thomas Kwasnitza, Dominik Stöckinger},
	pdfkeywords = {Higgs, Mass}
}

\allowdisplaybreaks

\newcommand{\GeV}{\text{GeV}}
\newcommand{\appref}[1]{Appendix~\ref{#1}}
\newcommand{\feft}{\texttt{Flex\-ib\-le\-EFT\-Higgs}\@\xspace}

\newcommand{\msf}[1]{\ensuremath{m_{\tilde{#1}_3}}}
\newcommand{\msq}{\msf{q}}
\newcommand{\msu}{\msf{u}}
\newcommand{\msd}{\msf{d}}
\newcommand{\ol}[1]{\overline{#1}}
\newcommand{\MSbar}{\ensuremath{\ol{\text{MS}}}\xspace}
\newcommand{\DRbar}{\ensuremath{\ol{\text{DR}}}\xspace}
\newcommand{\Lagr}{\mathcal{L}}
\newcommand{\ord}{\mathcal{O}}
\newcommand{\full}{\ensuremath{\text{full}}\xspace}
\newcommand{\eft}{\ensuremath{\text{eft}}\xspace}
\newcommand{\SM}{\ensuremath{\text{SM}}\xspace}
\newcommand{\MSSM}{\ensuremath{\text{MSSM}}\xspace}
\newcommand{\SUSY}{\ensuremath{\text{\sc SUSY}}\xspace}

\newcommand{\tL}{\ensuremath{0\ell}\xspace}
\newcommand{\oneL}{\ensuremath{1\ell}\xspace}
\newcommand{\twoL}{\ensuremath{2\ell}\xspace}
\newcommand{\thrL}{\ensuremath{3\ell}\xspace}
\newcommand{\fourL}{\ensuremath{4\ell}\xspace}
\newcommand{\nL}{\ensuremath{n\ell}\xspace}
\newcommand{\figref}[1]{figure~\ref{#1}}
\newcommand{\secref}[1]{section~\ref{#1}}
\newcommand{\tabref}[1]{table~\ref{#1}}

\newcommand{\tikzmark}[1]{%
	\tikz[overlay,remember picture, 
	baseline=-\the\dimexpr\fontdimen22\textfont2\relax]% correct vertical alignment:
	\node (#1) {};%
}

\makeatother

\begin{document}
	\maketitle
	\newpage

\section{Introduction}
\label{sec:1}

The method of effective field theory (EFT) is very useful to obtain
accurate descriptions of physics effects from heavy energy scales.
A
prominent example
is the calculation of the Higgs boson mass in supersymmetric (\SUSY)
models. Interestingly, the measured value ${M_h = (125.10 \pm 0.14 \,\GeV)}$
\cite{Aad:2015zhl,Tanabashi:2018oca} lies in the mass
range that can be accommodated by \SUSY models, however it typically
requires a rather heavy \SUSY particle spectrum. Hence many recent
precision calculations of the Higgs boson mass utilized
EFT methods
\cite{Draper:2013oza,Bagnaschi:2014rsa,Vega:2015fna,Lee:2015uza,
	Bagnaschi:2017xid,Braathen:2018htl,Gabelmann:2018axh,Allanach:2018fif,
	Harlander:2018yhj,Bagnaschi:2019esc,Kramer:2019fwz,Bahl:2019wzx}
 including hybrid methods which combine EFT and fixed-order calculations \cite{Hahn:2013ria,Bahl:2016brp,Athron:2016fuq, Staub:2017jnp,Athron:2017fvs,
 	Bahl:2017aev,Bahl:2018jom,R.:2019irs,Harlander:2019dge,
 	Bahl:2019hmm,Kwasnitza:2020wli,Bahl:2020tuq},
see ref.~\cite{Draper:2016pys,Slavich:2020zjv} for recent reviews.

In the present work we focus on the situation where the minimal
supersymmetric standard model (MSSM) is regarded as the full model
and matched to the standard model (SM) as the EFT. It corresponds to
assuming all \SUSY masses to be of the order of a common heavy scale $M_S$.
This situation is
the one of many Higgs boson mass calculations, but also of general
interest.

A crucial ingredient in EFT calculations is
the matching between SM and MSSM parameters.
The obtained threshold corrections or decoupling coefficients  stay finite in
the limit $M_S\rightarrow \infty$, but they are 
in general complicated functions of all dimensionless and dimensionful
parameters of the full model.
The MSSM contains several parameters which can lead to significant
systematic enhancements of higher-order threshold corrections.
A well-known example is the ratio of the two Higgs vacuum expectation
values $\tan\beta$. Higher-order
threshold corrections to the bottom-quark Yukawa coupling contain
terms of the form $g_3^{2n}(\tan\beta)^n$, where $g_3$ is
the QCD gauge coupling. These corrections arise for all $n$; they can be
of order $100\%$ and 
could spoil the convergence of perturbation theory. However, it was
observed and analyzed in detail in
refs.~\cite{Carena:1999py,Guasch:2003cv,Hofer:2009xb} 
 that these corrections
can be resummed at all orders. The all-order resummation only requires
ingredients from one-loop calculations.
Further generalizations of the analyses were presented in \cite{Noth:2008tw,Noth:2010jy, Marchetti:2008hw}.

Here we consider similar enhancements by the trilinear squark-Higgs
couplings $x_q$ arising in the threshold corrections for Yukawa
couplings and the quartic Higgs coupling $\lambda$. The
$\tan\beta$-enhancement mentioned above is strongly related to the
$x_b$-enhancement arising in the bottom Yukawa coupling. Our analysis
covers this but focuses mainly on the $x_t$-enhancements arising in
$\lambda$, which have a high impact on MSSM Higgs boson mass
predictions. First constraints on $x_t$-enhancements and statements on
$x_t$-resummation have been mentioned and used in
ref.\ \cite{Kwasnitza:2020wli}. Here we give proofs of these and more
general statements.

As discussed in ref.\ \cite{Kwasnitza:2020wli}, the computation of
threshold corrections allows the choice of different parametrizations:
Traditionally, EFT-parametrization is most common. Here the threshold
corrections are expressed as a perturbative expansion in terms of EFT
parameters (in practice truncated at some finite order). In contrast,
full-model parametrization means expansion in terms of full-model
parameters (and truncated at some finite order). In principle both
parametrizations are possible and equivalent, however after truncation
at finite order they differ. The behavior of enhanced
$x_q$-corrections depends crucially on the chosen parametrization, and
the resummation can be understood via comparing full-model and EFT
parametrization.

To provide a preview we compare the following three results for
contributions leading in the QCD gauge coupling and $x_t$ (valid at
sufficiently large $n$):
\begin{itemize}
	\item  diagrammatic contributions to the four-point Green function 
	  in the MSSM in full-model parametrization
          involve at most the following powers of $x_t$ (result from \secref{sec:GF})
	\begin{align}
	\label{eq:DGreen_MS}
	   	   	   \Gamma_{h^4}\supset (y_t^\MSSM)^4(g_3^\MSSM)^{2n}~ x_t^{\leq n}
	\end{align}
	\item threshold corrections to the quartic Higgs coupling in
          full-model parametrization involve at most (see
          ref.\ \cite{Kwasnitza:2020wli} and \secref{sec:threshold})
	\begin{align}
	\label{eq:DLambda_MS}
				\Delta \lambda \supset (y_t^\MSSM)^4(g_3^\MSSM)^{2n}~ x_t^{\leq 4}
	\end{align}
	\item threshold corrections to the quartic Higgs coupling in
          EFT-parametrization   involve at most (see 
          ref.\ \cite{Kwasnitza:2020wli} and \secref{sec:resummation}) 
		\begin{align}
		\label{eq:DLambda_sm}
		\Delta \lambda \supset (y_t^\SM)^4(g_3^\SM)^{2n}~x_t^{\leq (n +4 )}.
	\end{align}
\end{itemize}
Here the full-model (and EFT) parameters are denoted by a superscript
$\MSSM$ (and $\SM$) and are renormalized in the $\DRbar$ (and
$\MSbar$) scheme, respectively. We see that the results for threshold
corrections in full-model parametrization are the strongest; the
leading terms in EFT-parametrization can be resummed at all orders by
fixed-order calculations in full-model parametrization. These and more
general statements are the content of the present paper.

In \secref{sec:threshold} we derive the strong bound on the threshold corrections in parameters of the full model, similar to eq.~\eqref{eq:DLambda_MS}. We continue with the  weaker constraint for Green functions in \secref{sec:GF}, as in eq.~\eqref{eq:DGreen_MS}. Furthermore, we elaborate
how contributions to $\Gamma_{h^4}$ with highest powers in the $x_q$
cancel  in the matching procedure and thus how
eqs.\ \eqref{eq:DLambda_MS}, \eqref{eq:DGreen_MS} are compatible.
The discussion in \secref{sec:resummation}
gives details about the consequences of the constraint 
out of \secref{sec:threshold}.
It is outlined
what threshold corrections are required to
``resum'' the leading $x_q$ contributions at various
coupling structures. We continue to present predictions for the highest power $x_t$ contributions, as in eq.~\eqref{eq:DLambda_sm}, to threshold corrections at multi-loop level
and show the limitations.

\section{Constraints on threshold corrections $\Delta\lambda$ and $\Delta y$}
\label{sec:threshold}
As mentioned in \secref{sec:1}, we begin by discussing constraints on threshold corrections.
In our set-up the high-scale 
model is fixed to be the real MSSM in the same notation as in ref.~\cite{Kwasnitza:2020wli}. 
That is, all parameters  of the MSSM are denoted without a hat,
$P^{\MSSM} \equiv P$; most important are the MSSM gauge couplings
$g_1,g_2,g_3$, the third generation top and bottom Yukawa couplings
$y_q\in\{y_t,y_b\}$, the two Higgs vacuum expectation 
values $v_u, v_d ,$ $ {v=(v_u^2 +v_d^2)^{1/2}}$ and the ratio $\tan \beta =v_u/v_d$.
All MSSM parameters  are defined in the \DRbar scheme.

We consider the masses of all BSM fields to be close to one
characteristic scale $M_S$, which we assume to be much higher than the
electroweak scale.
The central interactions of our analysis involve two squarks $\tilde q\in \{\tilde t_L, \tilde t_R, \tilde b_L, \tilde b_R\}$ (interaction eigenstates) and one
Higgs scalar $\mathcal \phi \in \{h, H ,A,H^\pm, G^0, G^\pm\}$ (mass eigenstates). 
The corresponding trilinear couplings are dimensionful. Throughout the
present paper we assume these couplings to be large, i.e.\ of the
order $M_S$.

The Higgs sector mass eigenstates are obtained from the two Higgs
doublet components by unitary matrices depending on $\beta$ and
$\alpha$ at tree level. In the limit of
high \SUSY masses, $M_S\gg v$, the two
mixing angles  are related as $\alpha \approx \beta -\pi/2 $, and
the real trilinear tree-level interactions can be expressed as
\begin{align}
\label{eq:trilinear_sim}
\begin{split}
\Lagr^\MSSM=& y_t s_\beta  X_t  \,G^+ \tilde t^\dagger_R  \tilde b_L   
-\frac{iy_t c_\beta Y_t }{\sqrt{2}} \,A \tilde t^\dagger_R  \tilde t_L 
- \frac{ y_b c_\beta X_b}{\sqrt{2}}   \,h \tilde b^\dagger_R  \tilde b_L  
+ y_b s_\beta Y_b\,H^- \tilde b^\dagger_R   \tilde t_L  +\cdots,
\end{split}
\end{align}
and 20 other analogous terms. Neglecting powers of $i$ and $\sqrt{2}$,
each coupling of the relevant 24 operators $\phi \, \tilde q_R^{\phantom{\prime}}\tilde q^\prime_L$ is given by one of four parameter combinations 
\begin{align}
\label{eq:trilinear_beta}
\quad y_t s_\beta X_t,\qquad y_t s_\beta Y_t \cot{\beta} ,\qquad  &y_b
c_\beta X_b,\qquad y_b c_\beta Y_b\tan \beta
,
\end{align}
where we explicitly factored out the quark Yukawa combinations ($y_t
s_\beta$) and ($y_b c_\beta$), which correspond to the tree-level SM Yukawa
couplings.  
In terms of fundamental MSSM parameters the coupling coefficients of
interest are  \cite{Espinosa:2000df} 
\begin{subequations}
	\label{eq:trilinear_exp}
\begin{align}
	X_t= A_t -\mu \cot \beta,&&X_b= A_b -\mu \tan \beta  ,\\* 
	Y_t=A_t+ \mu \tan\beta,&& Y_b=A_b +\mu \cot\beta,
\end{align}
\end{subequations}
where $A_t$ and $A_b$ are the trilinear couplings from the soft breaking, 
$\mu$ is the higgsino mass term from the superpotential.   
The so-called squark mixing parameter $X_q$  also appears  in the
off-diagonal element of the stop- and sbottom mass matrix 
\begin{align}
\label{eq:stop_mass_mat}
\mathsf{M}^2_t =
\begin{pmatrix}
m_t^2 + \msq^2 & m_t X_t \\
m_t X_t & m_t^2 + \msu^2
\end{pmatrix}, &&
\mathsf{M}^2_b =
\begin{pmatrix}
m_b^2 + \msq^2 & m_b X_b \\
m_b X_b & m_b^2 + \msd^2
\end{pmatrix},
\end{align}
where $\msq$, $\msu$ and $\msd$ are soft breaking mass terms, i.e.~of order $M_S$. Note that $D$-terms have been neglected in the matrices. 
The quark masses  are $m_q= y_q f_q(\beta) v/\sqrt{2}\in\{m_t,m_b\}$
with the notation
\begin{align}
	f_q(\beta) = \begin{dcases}
	s_\beta,& q=t\\
  c_\beta,& q=b
	\end{dcases}\,.
\end{align}
To keep the analysis transparent, we introduce the dimensionless
parameters $x_q$ which indicate the appearance of any of the trilinear
couplings in \eqref{eq:trilinear_beta},
\begin{align}
	\label{eq:trilinear_xq}
x_t\in \left\{\frac{X_t}{M_S}, \frac{ Y_t \cot \beta}{M_S}\right\}, &&x_b\in \left\{\frac{X_b}{M_S}, \frac{ Y_b \tan \beta }{M_S}\right\}
\end{align}
where the tree-level SM Yukawa coupling is split off, leading to
$\cot\beta$ and $\tan\beta$ factors in connection with $Y_{t,b}$.

Next, we consider the SM as the valid EFT below the scale $M_S$
where all parameters are $\MSbar$ renormalized. We continue to use the notation as in ref.~\cite{Kwasnitza:2020wli};
the parameters of the SM are denoted with a hat $P^{\SM} \equiv \hat P
$, most relevant are the quartic Higgs coupling   $\hat \lambda$, the
SM gauge couplings $\hat g_1,\hat g_2,\hat g_3$,
the third generation quark-Yukawa couplings $\hat y_q \in \{\hat
y_t,\hat y_b\}$  and the Higgs VEV $ \hat v$.
In the SM the quark masses are given by $\hat m_q =\hat y_q \hat v/\sqrt{2} \in\{\hat m_t,\hat m_b\}$. 
All SM parameters are defined in the $\MSbar$ scheme.

As we consider a large mass gap $ M_S \gg \hat v$, the matching procedure results in a
relation between the parameters of both models which is expanded perturbatively.
The so-called threshold corrections can be written symbolically as $\Delta P =\hat P -P$.

The focus of the present section is on the appearance of the parametric enhancement $x_q^n$ in the threshold corrections of the Yukawa coupling $\Delta y_q$ and of the quartic coupling $\Delta \lambda$
\begin{align}
\label{eq:yuk_match}
\Delta y_q = &\hat y_q -f_q(\beta) y_q \\
\Delta \lambda=&\hat \lambda - \lambda
\end{align}
where the quartic coupling of the light Higgs $h$ in the MSSM is given
by $D$-terms as ${\lambda = (g_1^2 +g_2^2) c^2_{2\beta}/4}$.

The appearance  of the trilinear interaction via
eqs.~\eqref{eq:trilinear_beta} and \eqref{eq:stop_mass_mat} is always
accompanied with either a Yukawa coupling or a quark mass (which in
turn is proportional to a Yukawa coupling). This seems to suggest that
also in threshold corrections the power of $x_q$ is directly bounded
by the power of the Yukawa couplings. However, there are two
complications which might invalidate such a conclusion.
\begin{itemize}
\item
  The threshold correction in eq.~\eqref{eq:DLambda_sm} contains terms
  of the form $\hat g_3^2 \hat y_t^4 x_t^5$, which seem to violate the
  above conclusion. The origin of such terms has been analyzed in
  ref.~\cite{Kwasnitza:2020wli} and traced back to the implicit expansion of
  full-model parameters in terms of 
  EFT parameters. 
  In order to reveal the full $x_q$ dependence at $n$-loop,
  besides genuine $n$-loop diagrams also all
  such implicit
  expansions have to be inspected carefully.

\item
  The appearance of $m_q x_q$ in the squark mass matrices might be
  accompanied with factors of $1/m_q$ from loop integrations. One
  could proceed in two equivalent ways to extract the full $x_q$
  structure in this context:
	\begin{itemize}
		\item Transit to squark mass eigenstates and expand the 
		multi-loop Feynman integrals in powers of $x_q v/M_S$.
		\item Work in the chiral basis and treat the off-diagonal element as a
		two-squark vertex in Feynman graphs, which can be inserted arbitrarily often.
		We denote such insertions as chiral squark flips. Our analysis follows this approach.
	\end{itemize}
\end{itemize}

\subsection{Constraints}
\label{sec:constraint}
In this section we list the constraints on threshold corrections in full-model
parametrization. For a detailed discussion on the parametrization see sec.~2 of ref.~\cite{Kwasnitza:2020wli}.
In short, threshold corrections have to be expanded in a power series
in either EFT or
full-model parameters. 
Both options are equivalent; however, if truncated at finite order ($n$-loop) a difference of higher order ($>n$-loop) remains.
An important insight is that in full-model parametrization stronger bounds
for the powers in $x_q$ exist, as can be directly seen by comparing 	eqs.~\eqref{eq:DLambda_MS} and \eqref{eq:DLambda_sm}. 

Similar discussions on constraints can be found in the literature for $\Delta y_b$ , see ref.~\cite{Carena:1999py,Guasch:2003cv},  and  for $\Delta \lambda$, see ref.~\cite{Kwasnitza:2020wli}.
In the latter reference, this fact has already been used to achieve an 
$x_t$-resummation for the Higgs boson mass.

Here we present generalized constraints, which include finite powers of the Yukawa couplings
and which allow arbitrarily high orders in $\alpha_s$.
Furthermore, we consider gauge-less limit for
 $\Delta y_q$ but we allow a finite power of electroweak couplings in
 $\Delta \lambda$.

\begin{center}
	\begin{minipage}[c]{1\textwidth}
	  \textit{i) Consider the threshold correction $\Delta y_q$
            in full-model parametrization. In its leading- and
            subleading-QCD contributions of  $\ord( y_q g_3^{2n}+{y_q^3}g_3^{2n} )$ for any $n\ge 0$,
			the power of $x_q$ is at most the power of the
                        Yukawa coupling.
			Technically, we write the correction at these
                        orders as
			\begin{align}
		\label{eq:constraint_Delta_y}
			\Delta {y_q} 
			&= g_3^{2n} \, \left( {y_q}P_{1,n}(x_q) + y_q^3
			\, P_{3,n}(x_q) \right)+\ord(m/M_S)\,.
			\end{align}
			The coefficients are polynomials in $x_q$ with
                        indices corresponding to the coupling
                        structures. The desired constraints are
                        formulated as constraints on the degrees
                        of these polynomials:
		\begin{subequations}
			\begin{align}
						\label{eq:constraint_a1}
			\deg(P_{1,n})  \leq 1,\\
			\deg(P_{3,n}) \leq3.
			\end{align}
		\end{subequations}
		}
	\end{minipage}
\end{center}
\begin{center}
	\begin{minipage}[c]{1\textwidth}
		\textit{ii) Consider the  threshold correction
			$\Delta \lambda$ in full-model
                  parametrization. In its leading- and subleading-QCD
                  contributions of $\ord( y_q^4 g_3^{2n}+g_{1}^2
                  y_q^2g_3^{2n}+g_2^2 y_q^2 g_3^{2n}+ 
			  y_q^6 g_3^{2n})$ 
			 for any $n\ge 0$,
			 the power of $x_q$ is at most the power of the Yukawa coupling.
			 Technically, we write the correction at these
                         orders as
			\begin{align}
			\label{eq:constraint_Delta_lam}
			\Delta \lambda &= g_3^{2n} \, \left(y_q^4 P_{4,n}(x_q) + y_q^2 g_{1}^2  P_{2,n,g_1}(x_q)
			+  y_q^2 g_{2}^2  P_{2,n,g_2}(x_q)
			+  y_q^6 P_{6,n}(x_q)\right) +\ord(m/M_S)\,. 
			\end{align}
			The coefficients are polynomials in $x_q$,
                        with indices corresponding to the coupling structures,
                        whose degrees are constrained as
                        \begin{subequations}
                          	\begin{align}
				\label{eq:constraint_cy}
                                \deg(P_{4,n})  &\leq 4,\\
			  \label{eq:constraint_c1}\deg(P_{2,n,g_1})
                          &\leq 2,\\
				\label{eq:constraint_c2}\deg(P_{2,n,g_2})
                                &\leq 2,\\
				\deg(P_{6,n}) &\leq 6.
				\end{align}
			\end{subequations} 		
		}
	\end{minipage}
\end{center}

\subsection{Proof}
\label{sec:proof}

In the following we prove the above constraints by performing a
matching calculation between the MSSM and the SM valid below $Q=M_S$
with a focus on the third generation in the matter content.

The interactions of the MSSM are described by the \DRbar renormalized fields  e.g. $ \Phi  \in \{ h,   q,  g^{\mu}\}$ (the light Higgs boson, left- and right-handed top quark, left- and right-handed bottom quark, gluon)\footnote{We neglect the decomposition of $q$ in a left- and right-handed part, which later affects the decomposition of the wave function renormalization and the vertex function.} . Their kinetic term
is is canonically normalized.

For energy scales much below $M_S$ the effective theory can be described
through the SM-Lagrangian.
This can be expressed either by using
fields in terms of which renormalized Green functions
have the same normalization, or by using SM fields ${\hat \Phi  \in \{\hat h, \hat q, \hat g^{\mu}\}}$ (Higgs boson, left- and right-handed top, left- and right-handed bottom, gluon) which
are defined in the $\MSbar$-scheme and whose corresponding
kinetic term is canonically normalized.
In the following we will denote the relation between these differently
normalized fields as the wave function renormalization (WFR)
	\begin{align}
	\label{eq:WFR_K}
K_{\Phi}\equiv\frac{\delta \hat \Phi}{\delta \Phi}. 
\end{align}
We choose to neglect higher dimensional operators and
include dim$\,\leq4$ operators only.
Hence, the threshold corrections are evaluated in the limit  $v/M_S \rightarrow 0$. We make use of the advantage to set the Higgs VEVs to zero,
$v,v_{u,d} \rightarrow 0$, on the Lagrangian level, which
is equivalent to matching in the unbroken phase.
The Green functions $\Gamma$ obtained by this procedure are denoted as
\begin{align}
\label{eq:limitGreen}
\Gamma_{v=0} \equiv\tilde \Gamma.
\end{align}
The limit is carried out such that dimensionless parameter and couplings are fixed and all SM fields become massless.
Later in \secref{sec:GF}, we discuss the contributions to $\Gamma$ obtained by taking the limit $v\rightarrow0$ after
its evaluation in the broken phase.
The limit does not interchange with loop integration for IR sensitive loop functions
which allows for terms with higher powers in $x_q$.

As usual, the KLN theorem \cite{Kinoshita:1962ur, Lee:1964is, Sterman:1976jh} ensures that various IR-divergent contributions cancel and well defined threshold corrections remain.
 
The matching procedure is constructed by the simplified Green functions from eq.~\eqref{eq:limitGreen} and the 
corrections are determined at vanishing external momentum. 
In general, the matching conditions form a coupled system
in an evaluation at multi-loop level.
Since we are interested in threshold corrections to the Yukawa coupling ${y_q} $ and the quartic $\lambda$, 
the relevant subset of matching conditions is 

\begin{subequations}
	\label{eq:toy_matching}
\begin{align}
  \label{eq:yuk-match}
  \tilde \Gamma_{\bar qq  h}^{\full} &= \tilde \Gamma_{\bar qq h}^{\eft}, &&\rightarrow\hat y_q ,
\\
  \tilde \Gamma_{\bar qq  g^\mu}^{\full} &= \tilde \Gamma_{\bar qq g^\mu}^{\eft},
  &&\rightarrow \hat g_3, \\
  \tilde \Gamma_{ h^4}^{\full} &= \tilde \Gamma_{h^4}^{\eft}, &&\rightarrow \hat \lambda,	
		\label{eq:K_f}\\
\frac{\partial}{\partial \slashed p}	\left	[\tilde \Gamma_{\bar qq}^{\full}\right] &= \frac{\partial}{\partial \slashed p}	\left	[\tilde \Gamma_{\bar qq }^{\eft}\right], &&\rightarrow K_q,
	\\
	\label{eq:K_g}
\frac{\partial}{\partial p^2}	\left[	\tilde \Gamma_{ g^\mu g^\nu}^{\full} \right]
 &= \frac{\partial}{\partial p^2}	\left[	\tilde \Gamma_{ g^\mu g^\nu}^{\eft} \right],
 &&\rightarrow K_g, \\
 \frac{\partial}{\partial p^2}	\left[\tilde \Gamma_{ h^2}^{\full}\right] &= \frac{\partial}{\partial p^2}	\left[\tilde \Gamma_{h^2}^{\eft}\right], &&\rightarrow K_h.
	\end{align}
	\label{eq:toy_matching_WFR}
\end{subequations}
In the eqs.~\eqref{eq:toy_matching} the arrows associate symbolically
the matching condition with the corresponding parameter of the EFT. 
In order to study the appearances of $x_q$ in  $\Delta P$
at the multi-loop level, we inspect the following six kinds of loop contributions in the matching condition.

\begin{enumerate}[(i)]
	\item \textit{Genuine diagrammatic contributions to $\tilde \Gamma^{\full}$:} 
 The simplification of evaluating $\Gamma^\full$ in the unbroken phase implies the
	absence of bilinear chiral-mixing terms in internal squark propagators.
The only $x_q$ contributions arise from the 
trilinear vertices of two squarks and one Higgs component field $\phi \in\{h,H,A,H^\pm,G^0,G^\pm\}$.
As indicated by eq.~\eqref{eq:trilinear_beta}, each vertex 
carries an additional factor of the Yukawa coupling ${y_q}$.
Therefore, every external(internal) line $\phi$
is accommodated maximally by one(two) factor(s) of the Yukawa coupling.
The only Higgs interaction unrelated to
Yukawa couplings, which is considered in our discussion, is the quartic $D$-term interaction $\propto g_{1,2}^2 h^2 \tilde q^2$.

\item 
\textit{Counterterm insertions in diagrammatic contributions to $\tilde \Gamma^{\full}$:} 
In the \DRbar-scheme, counterterms which carry powers of $x_q$ are associated to divergent contribution in diagrams
which contain squark-Higgs interactions.
Hence, diagrams with counterterm insertions
can be analyzed as nested diagrammatic contributions as before.

\item
\textit{\DRbar$\rightarrow$ \MSbar conversion in  $\tilde \Gamma^{\full}$:} In order to transit from a \SUSY theory defined in the \DRbar scheme to a non-\SUSY theory in the \MSbar scheme one has to take unphysical $\epsilon$-scalars of mass $m_\epsilon$ in the SQCD sector into account. Their interactions are independent of the $x_q$ parameter. As outlined in ref.~\cite{Bednyakov:2007vm,Bednyakov:2009wt}, the scheme conversion 
can be performed by a decoupling of  $\epsilon$-scalars together with physical superpartners
at the threshold scale $M_S \approx m_\epsilon$. Further details have been given in refs.~\cite{Harlander:2007wh, Jack:1994rk, Martin:1993yx}.
	\item
	\textit{Conversion to canonically normalized fields in  $\tilde \Gamma^{\eft}$:}
	The matching condition in eq.\eqref{eq:toy_matching} needs to be combined with Green functions of canonically normalized
	fields.
	In order to extract unambiguously the WFR introduced in eq.~\eqref{eq:WFR_K} we write 
	\begin{align}
	\label{eq:wfr_con}
	\tilde \Gamma^{\eft}_{\Phi^n} = \left(K_{\Phi}\right)^n
	\tilde \Gamma^{\eft}_{\hat\Phi^n}.
	\end{align}
	\item 
	
		\textit{Diagrammatic contributions to $\tilde \Gamma^{\eft}$ (with canonically normalized fields):}
	We loop-expand the Green function in terms of 
	effective couplings of the EFT-Lagrangian 
	$P^{\eft} \in \{ \hat y_q, \hat \lambda, \hat g_{1,2},\hat{g}_3\}$
	\begin{align}
	\tilde \Gamma^{\eft}_{\hat\Phi^n}\equiv\left.\tilde \Gamma^{\eft} \right|_{\tL}
	+\left.\tilde \Gamma^{\eft} \right|_{\oneL} 
	+\left.\tilde \Gamma^{\eft} \right|_{\twoL} + \cdots\,.
	\end{align}
	\item 
	\textit{Expansion of $\tilde \Gamma^{\eft}$ in full-model parameters:}
	In order to have a consistent matching for the Wilson coefficient contained in the term
	$\left.\tilde \Gamma^{\eft} \right|_{\tL}$,  one has to evaluate
	and truncate
	both sides in eq.~\eqref{eq:toy_matching} in the same set of
	parameters. 
	In full-model parametrization the corrections 
	to $\left.\tilde \Gamma^{\eft} \right|_{m\ell} $ have to be expanded
	in terms of full-model parameters \cite{Kwasnitza:2020wli}
	\begin{align}
	\label{eq:double-loop_toy}
	\left.\tilde \Gamma^{\eft} \right|_{m\ell}=
	\left[\left.\tilde \Gamma^{\eft} \right|_{m\ell}\right]_{P^{\eft}=P^{\full}}	
	+\left[\frac{\partial}{\partial P^{\eft}}
	\left.\tilde \Gamma^{\eft} \right|_{m\ell}\right]_{P^{\eft}=P^{\full}}
	\Delta P + \cdots\,.	
	\end{align}
	This re-expansion in terms of renormalized parameters of $\Lagr^{\full}$ 
	mixes the loop orders of the perturbation theory such that a genuine contribution
	on the r.h.s.~of eq.~\eqref{eq:double-loop_toy} is of $(\geq m)$-loop.
	The joint evaluation 
	and truncation in one set of parameters 
	of $\tilde \Gamma^{\eft}$ is what we denote as double loop expansion. Note that, setting $v=0$ on the Lagrangian level results only in contributions to $\tilde \Gamma^\eft$ with finite powers in EFT parameters.\footnote{\label{footnotenonpolynomial}
Non-polynomial contributions like $\log (\hat m_q/Q)=\log \hat y_q +\cdots$ are absent in $\tilde \Gamma^\eft$.
  In the next chapter we will return to such contributions. }
\end{enumerate}
Now we trace all $x_q$ contributions entering $\Delta P$.
In the described steps, $x_q$ appears explicitly
in the items (i), (ii), (iv) and (vi). 
From the arguments in (i) and (ii) we conclude for $\tilde\Gamma^\full$
in the unbroken phase: indeed every factor $x_q$ originates from an 
interaction of squarks with Higgs bosons and hence is accompanied by
one factor of the coupling $y_q$.\footnote{Each internal line 
	of $\phi$ gives rise to two factors of ${y_q}$ whereas each external line induces one factor $y_q$.} Hence, each factor of $y_q$ may introduce
at most one factor of $x_q$ and we can write
\begin{subequations}
	\label{eq:Gamma_full}
	\begin{align} 
	\left. \tilde \Gamma_{h \bar qq}^{\full} \right. &\supset {y_q} g_3^{2n} x_q^{\leq 1}+ {y_q^3} g_3^{2n} x_q^{\leq 3},\\
	\left. \tilde \Gamma_{\bar qq  g^\mu}^{\full}\right.&\supset g_3  g_3^{2n} x_q^{\leq 0}+  y_q^2 g_3^{2n+1} x_q^{\leq 2},\\
	\left.  \tilde \Gamma_{h^4}^{\full}\right. &\supset y_q^2 g_{1,2}^2 g_3^{2n} x_q^{\leq 2} + y_q^4 g_3^{2n} x_q^{\leq 4}+ {y_q^6}  g_3^{2n} x_q^{\leq 6},\\
	\left.\frac{\partial}{\partial p^2}\tilde \Gamma^{\full}_{hh}\right.&\supset g_3^{2n} x_q^{\leq 0}+ y_q^2 g_3^{2n} x_q^{\leq 2},
	\\
	\left.\frac{\partial}{\partial p^2}\tilde \Gamma^{\full}_{g^\mu g^\mu}\right.&\supset g_3^{2n} x_q^{\leq 0}+ y_q^2 g_3^{2n} x_q^{\leq 2},
	\\
	\left.\frac{\partial}{\partial \slashed p}\tilde \Gamma^{\full}_{\bar qq}\right.&\supset g_3^{2n} x_q^{\leq 0} +  y_q^2 g_3^{2n} x_q^{\leq 2},
	\end{align}
\end{subequations}
where the first terms correspond to zero internal Higgs lines and the
last term corresponds to one internal Higgs line. 
The notation ``$\supset g_i^n x_q^{\leq m}$''
 means that at $\ord(g_i^n)$ (unsuppressed) contributions to the l.h.s.~with $x_q^{m+1}$ (or higher) are explicitly forbidden. 
However, the terms  on the l.h.s.~of ``$\supset $'' may contain 
other coupling structures  whose $x_q$ behavior is not specified.

The loop contributions to $\tilde \Gamma^{\eft}$  carry implicit $x_q$ contributions 
through items (iv) and (vi).  
Since $\tilde\Gamma^\eft$ at  $m$-loop  depends genuinely on the parameter shifts $\Delta P$ and on the WFR $K_{\Phi}$ at $(< m)$-loop, the evaluation of $\tilde \Gamma_{\Phi^n}^\eft$ is intertwined.
We continue by mathematical induction.
\begin{itemize}
	\item \textbf{Base step}:
	For the threshold corrections at the lowest order we  include the full-model contributions $\tilde \Gamma^{\full}$ at 1- and 2-loop analyzed in
	eq.~\eqref{eq:Gamma_full}. Performing a direct expansion of the $\tilde \Gamma^{\eft}$ in full-model parameters up to 2-loop results in
	\begin{subequations}
		\begin{align}
		\left.\Delta {y_q} \right. &\supset  {y_q}g_3^2 x_q^{\leq 1}+{y_q^3}  x_q^{\leq 3},\\
		\left.\Delta g_3 \right. &\supset g_3^3 x_q^{0} + g_3 y_q^2  x_q^{\leq 2},\\
		\left.\Delta \lambda \right. &\supset 
		g_{1,2}^2  y_q^2  x_q^{\leq 2}+ y_q^4 x_q^{\leq 4}+ y_q^6 x_q^{\leq 6},\\
		\left.K_\Phi\right.&\supset {g_3^{2}} x_q^{ 0} +{ y_q^2 } x_q^{\leq 2},
		\end{align}
	\end{subequations}
 for $\Phi \in\{q,g,h\}$ where  relevant orders only are taken into account.\footnote{Note that some contributions in $\Delta g_3$ and $K_\Phi$ are vanishing at 1-loop. But we list it to highlight the pattern.} 

	\item 	\textbf{Induction hypothesis:} We assume that the $\ord(g_3^{2n})$	corrections, i.e. $n$-, $(n+1)$- and $(n+2)$-loop, obtained
	from eqs.~\eqref{eq:toy_matching} satisfy 
	\begin{subequations}
		\label{eq:hypothes}
	\begin{align}
	\left.\Delta {y_q} \right. &\supset  {y_q}g_3^{2n} x_q^{\leq 1}+{y_q^3} g_3^{2n} x_q^{\leq 3},\\
	\left.\Delta g_3 \right. &\supset g_3^{2n+1} x_q^{0} + y_q^2 g_3^{2n+1} x_q^{\leq 2},\\
	\left.\Delta \lambda \right. &\supset g_{1,2}^2  y_q^2  g_3^{2n} x_q^{\leq 2}+ y_q^4 g_3^{2n} x_q^{\leq 4}+ y_q^6  g_3^{2n}x_q^{\leq 6},\\
	\left.K_\Phi\right.&\supset g_3^{2n} x_q^{ 0} +{ y_q^2 }g_3^{2n} x_q^{\leq 2}.
	\end{align}
\end{subequations}
	 Note that the 1-loop analysis is sufficient to show that $\Delta \lambda$ is of $\ord(g_{1,2}^2 y_q^2  g_3^{2n+2} x_q^{\leq2})$.\footnote{Because $\tilde \Gamma^{\eft}_{\hat h^4}$ has no contributions
	at $\ord( \hat g_{1,2}^2 {\hat y_q}^2 ,  \hat \lambda  {\hat y_q}^2)$ (and likewise for additional powers in $\hat g_3^2$), non-trivial contributions to $\tilde \Gamma^{\eft}_{\hat h^4}$ arise through the product $\lambda K_h$ only and not by reparametrization.}
	\item \textbf{Induction step}: Now we perform a matching at order $\ord(g_3^{2n+2})$.
	The full-model corrections at this order are characterized by
        eq.~\eqref{eq:Gamma_full} (with the replacement ${n\to n+1}$). 
	The genuine loop contributions to $\tilde \Gamma^{\eft}$ of
        item (v) above at
        the order $\ord(g_3^{2n+2})$  do not contain $x_q$ explicitly but need
        to be combined with $K_{\Phi}$  in eq.~\eqref{eq:wfr_con} and
	$\Delta P$ in
        eq.~\eqref{eq:double-loop_toy} up to $\ord(g_3^{2n})$ (where
        $\Phi \in\{q,g^\mu,h\}$ and $P \in\{{y_q},g_3,\lambda\}$).  
	Using the constraint from the  hypothesis
	eq.~\eqref{eq:hypothes}, the expansions in
        $\tilde\Gamma^\eft$ entering the evaluation of the matching
        equations \eqref{eq:toy_matching} can only lead to structures
        of higher order where each coupling $y_q$ is associated with
        at most one factor $x_q$.
	In conclusion
	the threshold correction can be evaluated to the same constraint as eq.~\eqref{eq:hypothes} with the substitution $g_3^{2n}\to g_3^{2n+2}$.
	This finishes the induction step and therefore the proof.
\end{itemize}

 \subsection{Comments}
 \label{sec:thresh_comments}
 \begin{itemize}
 	 	\item For phenomenological reasons we focused on low powers in $y_q$. However, the proof can be extended to orders with more Yukawa couplings.
 	 	\textit{In full-model parametrization, the highest power $x_q$ contributions to threshold corrections $\Delta \lambda$ 
 	 		and $\Delta y_q$ are of order $\ord(g_3^{2n} (y_t x_t )^{n_t} (y_b x_b )^{n_b})$}.
 	 	
 	\item The generalized constraint is in line with the statement of
          ref.~\cite{Noth:2008tw, Noth:2010jy} on resummation of
           mixed Yukawa corrections to $\Delta y_b$.
 In our notation the constraint is;
          \textit{ there are no $\ord(y_b (y_t^2 A_t \tan\beta/M_S)^n)$
          contributions to $\Delta y_b$}.
        
\item In the context of the MSSM including all three 
generations, the proof can be extended to soft-breaking 
trilinear couplings related to flavor violation. We will 
specify this statement in \secref{sec:application}.     
        
 	\item
 	Along the lines of \secref{proof} one could find
 	an alternative proof by analyzing the contributions
 	 in the large mass expansion.
In that context, eq.~\eqref{eq:Delta_v_taylor} provides the relevant
constraint on the leading QCD
 	contributions to quartic Higgs coupling and to the quark-Higgs coupling at $\ord( y_q^4  g_3^{2n})$ and $\ord({y_q} g_3^{2n})$, respectively.

\item
One may derive  similar constraints on Wilson coefficients of higher
dimensional operators, for instance on $H^6$. At one-loop order its threshold correction
contains terms $\propto y_q^6 x_q^6$.
By plugging in finite $v$, contributions to the $\ord(m/M_S)$ part of $\Delta\lambda$ in eq.~\eqref{eq:constraint_Delta_lam} would be generated. 
Those power-suppressed contributions 
involve higher $x_q$ powers than the ones given on the dimension-4
level by the constraint  on $\Delta \lambda$ in
eq.~\eqref{eq:constraint_Delta_lam}, which could 
also be studied by the arguments given in the present section. 
For a detailed analysis of such power-suppressed terms see ref.~\cite{Bagnaschi:2017xid}; the discussion involves the inspection of ``hedgehog diagrams''
 as in ref.~\cite{Hofer:2009xb}.

\end{itemize}

\section{Constraints on the Green functions}
\label{sec:GF}
In this section we present the constraint on Green functions
$\Gamma_{ h^4}$ and $\Gamma_{  h \bar qq}$ in the MSSM
which is weaker than in the case for threshold corrections.\footnote{%
  Here the Green functions are evaluated as usual with $v\ne0$,
  i.e.\ in the broken phase, in contrast the Green functions $\tilde
  \Gamma$ used in the proof 
  in \secref{sec:proof}.}
In the light of a large mass splitting between the SM fields 
and the \SUSY scale $m < M_S$, we analyze the structure of the Green function at 
leading order in the expansion parameter $m/M_S$.

\begin{center}
	\begin{minipage}[c]{0.9\textwidth}
		\textit{ i) In full-model parametrization unsuppressed 
			$x_q$ contributions to the 
			Green function $\Gamma_{\bar q q h}$ at $\ord(g_3^{2n})$ 
			and  $n\ge 1$ are expanded as
			\begin{align}
			\label{eq:constr_gamma_hqq}
			\left.\Gamma_{\bar q q h} (p=0)\right|_{{y_q}g_3^{2n}}
			&={y_q} g_3^{2n} \, \bar P_{1,n}(x_q) +\ord(m/M_S).
			\end{align}
			The coefficients are polynomials in $x_q$
                        whose degrees are constrained as  
			\begin{align}
			\deg(\bar P_{1,n}) \leq 
                        \begin{cases}
			1&\text{ for } n\leq 2,  
			\\  n-1&\text{ for } n \geq 3.  \end{cases}
			\label{eq:Gamma_hff_constraint}
			\end{align}			
		}
	\end{minipage}
\end{center}
\begin{center}
	\begin{minipage}[c]{0.9\textwidth}
		\textit{ ii) In full-model parametrization unsuppressed 
			$x_q$ contributions to the 
			Green function $\Gamma_{h^4}$ at $\ord(  y_q^4g_3^{2n})$ 
			and  $n\ge 1$ are expanded as
			\begin{align}
			\label{eq:constr_gamma_hhhh}
			\left.\Gamma_{h^4} (p=0)\right|_{
                          y_q^4g_3^{2n}} &= y_q^4 g_3^{2n} \bar P_{4,n}(x_q) +\ord(m/M_S),
			\end{align}
			where $\bar P_{4,n}$ are polynomials whose
                        degrees are
                        constrained as
			\begin{align}
			\deg(\bar P_{4,n})\leq 
			 \begin{cases}
			4\text{ for } n\leq4,
			\\ n\text{ for } n\geq5.
			\label{eq:Gamma_h^4_constraint}
			 \end{cases}
                        \end{align}			
		}
	\end{minipage}
\end{center}

To exemplify the $x_q$ contributions which are allowed by the theorem
we show  \tabref{tab:gamma}. The constraints on Green functions in
eqs.\ (\ref{eq:Gamma_hff_constraint},\ref{eq:Gamma_h^4_constraint})
can be compared to the constraints on threshold corrections in
eqs.\ (\ref{eq:constraint_a1},\ref{eq:constraint_cy}), respectively.
The constraints on Green functions are weaker and allow terms with
higher power of $x_q$  than  of ${y_q}$. 

\begin{table}[h]
	\centering
	\begin{tabular}{ccc}
		\toprule
  loop order & $\Gamma_{h^4}$ & $\Gamma_{\bar q q h}$ \\
\midrule
\oneL & $  y_q^4  x_q^{\leq4} $ & $ {y_q} g_3^2  x_q^{\leq 1} $\\[0.5em]
\twoL & $  y_q^4  g_3^2 x_q^{\leq4} $ & $ {y_q} g_3^4  x_q^{\leq 1} $\\[0.5em]
\thrL & $  y_q^4  g_3^4 x_q^{\leq4} $ & $ {y_q}  g_3^6  x_q^{\leq 2}$\\[0.5em]
\fourL & $  y_q^4  g_3^6 x_q^{\leq4} $ & $ {y_q}  g_3^8   x_q^{\leq 3}$\\[0.5em]
$5\ell$ & $  y_q^4  g_3^8 x_q^{\leq4} $ & $ {y_q}  g_3^{10}   x_q^{\leq 4}$\\[0.5em]
$6\ell$ & $  y_q^4  g_3^{10} x_q^{\leq5} $ & $ {y_q}  g_3^{12}   x_q^{\leq 5}$\\[0.5em]
$7\ell$ & $  y_q^4  g_3^{12} x_q^{\leq6} $ & $ {y_q}  g_3^{14}   x_q^{\leq 6}$\\[0.5em]
\vdots & \vdots & \vdots \\
		\bottomrule
	\end{tabular}
	\caption{Contributions to $\Gamma_{\bar q q h}^{\nL}$ and $\Gamma_{h^4}^{\nL}$, 
		which are not constrained to be vanishing by the theorem.}
	\label{tab:gamma}
\end{table}

\subsection{Proof}
\label{proof}
The proof of the theorem relies on an inspection of diagrammatic contributions
which is based on a large mass expansion (LME).
The strategy is to analyze a genuine multi-loop diagram and track all $x_q$ appearances as described in the following.

The evaluation of an arbitrary diagrammatic contribution 
	$F$ to a Green function with the technique of LME
	leads	to a sum of diagrams $G_g\equiv F /g$ with insertions $\Delta V_{g}$
	stemming from the subgraph $g$
	\begin{align}
	\label{eq:LME_general}
	F &= \sum_{g} G_g \circ \Delta V_g.
	\end{align}
	More specifically $\Delta V_g$ represents the Taylor expansion 
	of the subdiagram $g$ which is 1PI in the light lines and contains 
	all heavy internal lines. 

	In order to investigate the contributions to $\Delta V_g$	it is of advantage to introduce the notion of an \textit{effective vertex}, which makes the factorization
	property of $\Delta V_g$ manifest.
	That is, if $g$ is an unconnected set of $k$ connected graphs, 
	the result of its Taylor expansion  factorizes symbolically as 
	$\Delta V_g \supset \Delta v_1\times \cdots \times \Delta v_k$.
	
	\paragraph{	\underline{Effective vertex	$\Delta v_i$}}
	 In the LME of eq.~\eqref{eq:LME_general}, the effective vertex $\Delta v_i$ denotes the contraction of a connected subdiagram in $g$. This contraction represents	the point-like connection of light lines, which originates from a Taylor expansion
	 in the light mass of the in- or outgoing fields and their soft momentum.
	 This procedure can be conceived as a construction of a BRST invariant
	 EFT below the scale $M_S$, where $\Delta v_i \equiv \Delta v_{\mathcal{O}}$ 
	 is correlated  to a threshold
	 correction of  a coupling corresponding to an
	 operator $\mathcal O \subset \mathcal L_{\text{EFT}}$ \cite{Beneke:1997zp, Smirnov:2002pj}.\footnote{In order to extract the threshold correction, it is mandatory to substitute the fields by ones which are canonically normalized and the WFR \eqref{eq:WFR_K}.}

	In \appref{app:1} we analyze all $\Delta v_i$  relevant for 
$\Gamma_{\bar q q h}$ and $\Gamma_{h^4}$ at leading order in the QCD coupling.
	After truncating the Taylor expansion at leading order in $m/M_S$ 
the result can be expressed as 	
\begin{align}
\Delta v_{\mathcal O} = g_3^{2n} P(x_q),
\end{align}
where the degree of the polynomial in $x_q$ inside of r.h.s.~depends 
on the fields $\phi \in \{h,t,g,c\}$  which are
attached to the individual effective vertex.
More specifically, in  \appref{app:1} we derive that the maximal
$x_q$ contribution to an effective vertex 
can be written as
\begin{align}
\label{eq:Delta_v}
\Delta v_{\mathcal{O}}\supset 
& ({y_q} x_q)^{n_h} (x_q m)^{n_{d} -n_p} p^{n_p}
\left(x_q\frac{m}{M_S}\right)^{n_{I}+n_p-n_{d} }
\left[\text{const} + \ord\left( \frac{m,p}{M_S}  \right)\right],
\end{align}
where the indices $n_h$, $n_d$, $n_p$ and $n_I$ represent the number of
Higgs lines attached by the vertex ($n_h$), the mass dimension ($n_d$),
the power of momentum ($n_p$) and the number of chiral flips in the squark propagators ($n_I$). 
We note that eq.~\eqref{eq:Delta_v} confirms the constraints
presented in \secref{sec:constraint}, where the threshold corrections
have been expanded in full-model parameters.  
	For our purpose, we classify the effective vertices for our discussion as:
	\begin{itemize}
		\item An infinite number of vertices with a suppression by the \SUSY scale, which correspond to non-renormalizable
		operator in the EFT: $\Delta v \propto 1/M_S ^{\geq1}$
                (note that because of BRST invariance, trilinear interactions of a Higgs with gluons/ghosts 
                correspond to higher dimensional operators)
                \footnote{We note that higher dimensional operators
		can induce an $x_q$ dependence of higher orders, see ref.~\cite{Bagnaschi:2017xid}.}
		\begin{align}
		\label{eq:higher_dim}
			\Delta v_{hgg}, &&\Delta v_{h\bar cc},&&\Delta v_{q^2h^2},&&\cdots.
		\end{align}
		\item A finite number vertices which correspond to  renormalizable operators which,
		however, have no explicit $x_q$ dependence, $\Delta v \propto g_3^{2n} x_q^0$:
		\begin{align}
				\label{eq:effvert_lowdim}
		  \Delta v_{gg}, &&\Delta v_{g^3},&&\Delta v_{g^4}, &&\Delta v_{g\bar cc}, &&\Delta v_{\bar cc}, &&\Delta v_{g\bar qq}, &&\slashed p \text{~part of~}\Delta v_{\bar qq}, &&\cdots,
		\end{align}
		where the dots represent interactions which do not contribute at leading orders
		in $ g_3^{2n}$. In Feynman diagrams, those are denoted by a white square.
		\item The remaining list of effective vertices corresponds to renormalizable
		operators which introduce an explicit $x_q$ dependence. These are
		the three vertices
		\begin{align}
		\label{eq:list}
					  \Delta v_{\bar qqh}\propto y_q g_3^{2n} x_q^{\leq 1}, && m_q \mathbbm{1} \text{~part of~}\Delta v_{\bar qq}\propto g_3^{2n} x_q^{\leq 1}, &&\Delta v_{h^4}\propto g_3^{2n} x_q^{\leq 4}.
		\end{align}
		In the diagrams, those effective vertices are denoted by a gray square.
	\end{itemize}

	Now we apply dimensional analysis to evaluate
	each contribution in eq.~\eqref{eq:LME_general} 
	which depends on $\Delta v_i$ and a loop integral  function $I$
	originating from the integral over loop momenta in $G_g$ 
	\begin{align}
	G_g\circ\Delta V_l= \Delta v_1^{j_1}  \Delta v_2^{j_2}\cdots\Delta v_k^{j_k}  I(m)+ \cdots\,.
	\end{align}
	Here each effective vertex is raised to some integer power
        $j_i$, and the only physical scale in the function $I$ is the 
	light mass $m$.
	Hence, the loop integral $I(m)$ cannot induce further $M_S$
	enhancement. In conclusion, the final $x_q$ dependence and 
	the suppression of $\ord(m/M_S)$ in each addend in eq.~\eqref{eq:LME_general} 
	can be read off the product of effective vertices.
        Once a certain loop order and power of the couplings ${y_q}$ and $ g_3^2$ is
        specified, it is a matter of inspecting all possible
        contributions to construct diagrams with the vertices $\Delta v_i$ which maximize the 
	power of $x_q$ appearing at this order. In the following we
        present the results of this analysis.
	\begin{itemize}
		\item $\Gamma_{\bar q q h}$: Depending on the loop level
		the maximal $x_q$ powers originates from contributions
    in eq.~\eqref{eq:LME_general}  of the form
		\begin{align}
		\ord({y_q} ({g_3^2})^{n\leq 2}):&&
		\begin{tikzpicture}[baseline=(c)]
		\begin{scope} [shift={(0,0)}]	
		\node (c) at (0,-0.1){};
		\node at (.4,.3) {$q$};	
\node at (1.75,.3) {$q$};
\node at (1.4,1.) {$h$};
		\draw[fermion] (0,0)--(1,0);
		\draw[fermion] (1.2,0)--(2.2,0);
		\draw[scalarnoarrow] (1.1,1.1)--(1.1,0.1);
				\begin{scope}[shift={(0,0)}]
		\draw[fill=gray](1.,-0.1) rectangle (1.2,0.1);
		\end{scope}
		\end{scope}
		\end{tikzpicture} &\propto \Delta v_{\bar qqh} 
		&&\supset {y_q} g_3^{2n}  x_q^{\leq 1},\\
		\ord({y_q} ({g_3^2})^{n>2}):&&
		\begin{tikzpicture}[baseline=(c)]
		\begin{scope} [shift={(0,0)}]	
		\node (c) at (0,-0.1){};
		\node at (-.3,.3) {$q$};	
		\node at (2.5,.3) {$q$};
		\node at (2.1,1.) {$h$};
		\draw[fermionnoarrow] (-0.5,0)--(2.7,0);
		\draw[fermion] (2.35,0)--(2.7,0);
		\draw[fermion] (-0.5,0)--(0.1,0);
		\draw[scalarnoarrow] (1.85,1.1)--(1.85,0.1);
		\draw[gluon] (2.11,0) arc (0:-180:1.045);
		\begin{scope}[shift={(.75,0.)}]
\draw[fill=gray](1.,-0.1) rectangle (1.2,0.1);
\end{scope}	
	\begin{scope}[shift={(.35,0.)}]
\draw[fill=gray](1.,-0.1) rectangle (1.2,0.1);
\end{scope}		
\begin{scope}[shift={(-.4,0.)}]
\draw[fill=gray](1.,-0.1) rectangle (1.2,0.1);
\end{scope}
\begin{scope}[shift={(-.8,0.)}]
\draw[fill=gray](1.,-0.1) rectangle (1.2,0.1);
\end{scope}
		\begin{scope}[shift={(1.1,.1)}, scale=0.7, every node/.append style={transform shape}]
		\node at (0,0) {$\cdots$};	
		\end{scope}
		\end{scope}
		\end{tikzpicture}& \propto  g_3^2 \Delta v_{\bar qqh}  \left(\Delta v_{\bar qq}\right)^{n-2}
		&&\supset {y_q} g_3^{2n}  x_q^{\leq (n-1)},
		\end{align} 
		which establishes the constraint out of \eqref{eq:Gamma_h^4_constraint}.
		
				\item $\Gamma_{h^4}$: Depending on the loop level
		the maximal $x_q$ powers originates from contributions
		in eq.~\eqref{eq:LME_general}  of the form
		\begin{align}
		\ord( y_q^3 ({g_3^2})^{n\leq 4}):&&
\begin{tikzpicture}[baseline=(c)]
\begin{scope} [shift={(0.,0)}]
\draw[scalarnoarrow](145:1.5) -- (145:.15cm);
\node at (145:1.65) {$h$};
\draw[scalarnoarrow](215:1.5) -- (215:.15cm);
\node at (215:1.7) {$h$};
\draw[scalarnoarrow](35:1.5) -- (35:.15cm);
\node at (35:1.65) {$h$};
\draw[scalarnoarrow](-35:1.5) -- (-35:.15cm);
\node at (-35:1.7) {$h$};
\begin{scope}[shift={(-1.1,0)}]
\draw[fill=gray](1.,-0.1) rectangle (1.2,0.1);
\end{scope}
\end{scope}
\end{tikzpicture} &\propto \Delta v_{h^4} 
		&&\supset  y_q^4 g_3^{2n}  x_q^{\leq 4},
		\\
		\label{eq:gamma_h4_inserions}
		\ord( y_q^4 ({g_3^2})^{n>4}):&&
	\begin{tikzpicture}[baseline=(c)]
	%external legs
	\draw[scalarnoarrow](145:1.5) -- (145:.5cm);
	\node at (145:1.65) {$h$};
	\draw[scalarnoarrow](215:1.5) -- (215:.5cm);
	\node at (215:1.7) {$h$};
	\draw[scalarnoarrow](35:1.5) -- (35:.5cm);
	\node at (35:1.65) {$h$};
	\draw[scalarnoarrow](-35:1.5) -- (-35:.5cm);
	\node at (-35:1.7) {$h$};
	\node (c) at (0,-0.1){};
	%%inner ring
	\begin{scope} [shift={(-1.5,0)}]
	\begin{scope}[shift={(1.12,1.45)}, rotate=-75]
	\draw[fermionloop] (2,0) arc (360:0:.5);
	\end{scope}
%mass insertions
	  \begin{scope}[shift={(1.7,-0.45)}, rotate=30]
\draw[fill=gray](.,-0.1) rectangle (.2,0.1);
	  \end{scope}
	  \begin{scope}[shift={(1.1,-0.35)}, rotate=-30]
\draw[fill=gray](.,-0.1) rectangle (.2,0.1);
	 \end{scope}
	% squares
	\node (c) at (0,-0.1){};
	\end{scope}
			\begin{scope}[shift={(0,-0.6)}, scale=0.7, every node/.append style={transform shape}]
	\node at (0,0) {$\cdots$};	
	\end{scope}
	\end{tikzpicture}
	 &\propto  y_q^4 \left( \Delta v_{\bar qq}\right)^{n}
		&&\supset  y_q^4 g_3^{2n}  x_q^{\leq n},
		\end{align} 
				which establishes the constraint out of \eqref{eq:Gamma_hff_constraint}.
	\end{itemize}

\subsection{Explicit cancellation of contributions at $\ord( y_q^4g_3^{10} x_q^5)$}
\label{sec:expl_cancel}
In this section we elaborate the connection between the constraints on the Green
functions and on the threshold correction.
As a consistency check of both, we discuss various contributions in a
matching calculation of Green functions $\Gamma_{h^4}$ 
for the threshold correction $\Delta \lambda$. We choose to work at
the 6-loop order $\ord( y_q^4 g_3^{10})$, which provides an
instructive illustration.
In contrast to eq.~\eqref{eq:toy_matching_WFR} evaluated in the
unbroken phase, we now consider the matching condition for Green
functions at finite $v$ (but still at vanishing external momenta)
\begin{align}
\label{eq:matching_constraint}
\Gamma_{h^4}^{\full}= \Gamma_{h^4}^{\eft}.
\end{align}
The constraint in eq.~\eqref{eq:Gamma_h^4_constraint} allows  
 ${\Gamma_{h^4}^{\full}\supset  y_q^4g_3^{10} x_q^{\leq 5}}$,
i.e.\  $x_q^5$ may appear in this equation.
However, the stronger constraint in eq.~\eqref{eq:constraint_cy},
$\Delta \lambda \supset  y_q^4g_3^{10} x_q^{\leq 4}$, excludes explicitly
contributions $\propto x_q^5$ in the threshold correction. Our goal in
the following is to explain how these two statements are compatible.

It turns out that the main mechanism is a  cancellation of such $x_q^5$ contributions in
eq.~\eqref{eq:matching_constraint} due to an
interplay between a large mass expansion of
diagrams on the l.h.s.~and the double loop expansion on the 
r.h.s.~of eq.~\eqref{eq:matching_constraint} (see the discussion
around eq.~\eqref{eq:double-loop_toy} for details on the necessary
double loop expansion). We present 
our reasoning for both contributions in the following.

\paragraph{\underline{$\Gamma_{h^4}^{\full}$ at $\ord( y_q^4 g_3^{10} x_q^5)$}}
 		We consider the sum of 6-loop diagrams which could give rise
 		to a contribution of $\ord( y_q^4 g_3^{10} x_q^5)$
 		to the Green function in the full model.
 		From eq.~\eqref{eq:gamma_h4_inserions} we know that diagrammatic contributions with highest power in $x_q$ can be characterized under the LME as
\begin{align}
\label{eq:expand_Gamma_h4_SUSY}
\begin{split} i\Gamma_{h^4}^{\full}=
\sum_{\text{1PI diagrams}}
\begin{tikzpicture}[baseline=(c)]
\begin{scope} [shift={(0.,0)}]
\begin{scope} [shift={(-1.5,0)}]
\draw[fermionnoarrow] (1,0) arc (180:0:.5);
\draw[fermionnoarrow] (2,0) arc (0:-180:.5);
\node at (1.5,0){\scriptsize{1PI}};
\end{scope}
\draw[scalarnoarrow](145:1.5) -- (145:.5cm);
\node at (145:1.65) {$h$};
\draw[scalarnoarrow](215:1.5) -- (215:.5cm);
\node at (215:1.7) {$h$};
\draw[scalarnoarrow](35:1.5) -- (35:.5cm);
\node at (35:1.65) {$h$};
\draw[scalarnoarrow](-35:1.5) -- (-35:.5cm);
\node at (-35:1.7) {$h$};
\node (c) at (0,-0.1){};
\end{scope}	
\end{tikzpicture}
\overbrace{=}^{\text{LME}}&
\begin{tikzpicture}[baseline=(c)]
\begin{scope} [shift={(0.,0)}]
\draw[scalarnoarrow](145:1.5) -- (145:.15cm);
\node at (145:1.65) {$h$};
\draw[scalarnoarrow](215:1.5) -- (215:.15cm);
\node at (215:1.7) {$h$};
\draw[scalarnoarrow](35:1.5) -- (35:.15cm);
\node at (35:1.65) {$h$};
\draw[scalarnoarrow](-35:1.5) -- (-35:.15cm);
\node at (-35:1.7) {$h$};
\begin{scope}[shift={(-1.1,0)}]
\draw[fill=gray](1.,-0.1) rectangle (1.2,0.1);
\end{scope}
\end{scope}
\end{tikzpicture} (=F)\\
&+ 
\begin{tikzpicture}[baseline=(c)]
%external legs
\draw[scalarnoarrow](145:1.5) -- (145:.5cm);
\node at (145:1.65) {$h$};
\draw[scalarnoarrow](215:1.5) -- (215:.5cm);
\node at (215:1.7) {$h$};
\draw[scalarnoarrow](35:1.5) -- (35:.5cm);
\node at (35:1.65) {$h$};
\draw[scalarnoarrow](-35:1.5) -- (-35:.5cm);
\node at (-35:1.7) {$h$};
\node (c) at (0,-0.1){};
%%inner ring
\begin{scope} [shift={(-1.5,0)}]
  \begin{scope}[shift={(1.12,1.45)}, rotate=-75]
  \draw[fermionloop] (2,0) arc (360:0:.5);
  \end{scope}
    \begin{scope}[shift={(1.5,-0.5)}, scale=1]
  \draw[fill=gray]   (-0.1,-0.1) rectangle (0.1,0.1);
  \end{scope}
  \begin{scope}[shift={(1.92,0.27)} ,rotate=45, scale=1]
  \draw[fill=gray] (-0.1,-0.1) rectangle (0.1,0.1);
  \end{scope}
  \begin{scope}[shift={(1.92,-0.27)} ,rotate=45, scale=1]
  \draw[fill=gray] (-0.1,-0.1) rectangle (0.1,0.1);
  \end{scope}
  \begin{scope}[shift={(1.08,-0.27)} ,rotate=45, scale=1]
  \draw[fill=gray] (-0.1,-0.1) rectangle (0.1,0.1);
  \end{scope}
  \begin{scope}[shift={(1.08,0.27)} ,rotate=45, scale=1]
  \draw[fill=gray] (-0.1,-0.1) rectangle (0.1,0.1);
\end{scope}
\node (c) at (0,-0.1){};
\end{scope}
\end{tikzpicture} (=G)\\
&+ \cdots \,,
\end{split}
\end{align}
where we classified the highest $x_q$ powers in the resulting diagrams in the following way.
\begin{itemize}
	\item As discussed in eqs.~\eqref{eq:list} and \eqref{eq:LME_h^4}, diagrams of type $F$ with ``hard'' loop momenta  never give rise to unsuppressed higher
	$x_q$ contributions  than
	$\ord( y_q^4 g_3^{10} x_q^4 )$
		\begin{align}
	F\supset  y_q^4 g_3^{10} x_q^{4} \left(x_q \frac{m}{M_S}\right)^{n_I},
	\end{align}
	for some integer $n_I\geq0$.
	\item At $\ord( y_q^4g_3^{10})$ diagrams of type $G$ have one light
	``soft''
	loop momentum $k$ in the internal quark line. 
	They are the leading-power contribution in the $x_q$ parameter as noted
	in eq.~\eqref{eq:gamma_h4_inserions} and they can be
        schematically evaluated to 
\begin{subequations}
\begin{align}
G&= \Delta v_{ \bar qq} (\Delta v_{\bar qq h})^4 \int_{k} f(k,m_q)\\
  &\supset  y_q^4 g_3^{10} x_q^{\leq 5} m_q ~I(m_q,Q) ,
\end{align}
  \label{eq:G_contribution}
\end{subequations}
where $f(k,m_q)$ denotes the loop integrand, the effective vertices $\Delta v_{\bar qq h}$ and
 $\Delta v_{ \bar qq}$ carry each one factor of $\ord( g_3^2 x_q)$ at 1-loop,
 respectively. 
The crucial difference to the diagram $F$ is that
higher-power contributions as $x_q^{n}$, for $n>4$, do not lead to 
a suppression by $m/M_S$.
Note that the two-vertex in diagram $G$ acts like a mass insertion, 
which corresponds to the fact that $\Delta v_{\bar q q}|_{x_q}\propto m_q$ 
(in contrast $\Delta v_{ \bar qq h}$ is dimensionless).
As can be seen by eq.~\eqref{eq:list}, the  1-loop QCD enhanced contributions
to the effective vertices can be decomposed as
\begin{subequations}
\label{eq:1-loop-as}
\begin{align}
\Delta v_{\bar qq}&= -im_q  g_3^2  \, a_{(1,1)} x_q + \cdots,\\
\Delta v_{ \bar qq h}&= -i{y_q} f_q(\beta) g_3^2 \, a_{(1,1)} x_q + \cdots,
\end{align}
\end{subequations}
where the dots denote unsuppressed contributions with less powers in $x_q$ and $a_{(1,1)}$ is some coefficient independent of $\beta$.
For the diagram $G$ to be dimensionless implies that the
integral function $I(m_q,Q)$ in eq.~\eqref{eq:G_contribution}
has to be of negative mass dimension.
However, it cannot induce any suppression by the heavy scale $M_S$. It
only depends on light physical scales of order $m$, i.e.~$I(m_q,Q)\propto 1/m$.
In sharp contrast to eq.~\eqref{eq:Gamma_full} the Green function has higher power in $x_q$ contribution
\begin{align}
	 \Gamma^\full &\supset  y_q^4 g_3^{10} x_q^{\leq 5},
\end{align}
which stays finite in the limit $v\rightarrow 0$.
In consequence the only source of potential
contributions at $\ord( y_q^4 g_3^{10} x_q^5)$ can only stem from
diagrams of type $G$ and of similar types, where a combination of effective vertices,
 $(\Delta v_{\bar qq})^{n} (\Delta v_{\bar qqh})^{(5-n)} $  with $1\leq n\leq5$,
are connected by an internal fermion loop.
\end{itemize}

\paragraph{\underline{$\Gamma_{h^4}^{\eft}$ at $\ord( y_q^4 g_3^{10} x_q^5)$}}
In this paragraph, we discuss the EFT loop contributions to the 
r.h.s.~of eq.~\eqref{eq:matching_constraint}. 
In order to connect to the constraint on
the threshold correction $\Delta \lambda$ in full-model parameters,
we have to double loop expand the EFT Green function as
 outlined in eq.~\eqref{eq:double-loop_toy}. The relevant contributions originate from the
 1-loop diagram with an internal  quark $\left.\Gamma_{h^4}^{\eft}\right|_{\oneL}$.
We write the double loop expansion of this 1-loop
  contribution symbolically as
 \begin{align}
\label{eq:expand_Gamma_h4_SM}
\begin{split}
\left.i\Gamma_{h^4}^{\eft}\right|_{\hat y_q^4}
=
\left[
e^{\Delta {y_q}\frac{\partial}{\partial \hat y_q}}
\begin{tikzpicture}[baseline=(c)]
\begin{scope} [shift={(0.,0)}]
\begin{scope} [shift={(-1.5,0)}]
\draw[fermionloop] (1,0) arc (180:0:.5);
\draw[fermionloop] (2,0) arc (0:-180:.5);
\end{scope}
\draw[scalarnoarrow](145:1.5) -- (145:.5cm);
\node at (145:1.65) {$h$};
\draw[scalarnoarrow](215:1.5) -- (215:.5cm);
\node at (215:1.7) {$h$};
\draw[scalarnoarrow](35:1.5) -- (35:.5cm);
\node at (35:1.65) {$h$};
\draw[scalarnoarrow](-35:1.5) -- (-35:.5cm);
\node at (-35:1.7) {$h$};
\node (c) at (0,-0.1){};
\end{scope}	
\end{tikzpicture}
\right]_{\hat y_q = y_q f_q(\beta)}
=&
\begin{tikzpicture}[baseline=(c)]
%external legs
\draw[scalarnoarrow](145:1.5) -- (145:.5cm);
\node at (145:1.65) {$h$};
\draw[scalarnoarrow](215:1.5) -- (215:.5cm);
\node at (215:1.7) {$h$};
\draw[scalarnoarrow](35:1.5) -- (35:.5cm);
\node at (35:1.65) {$h$};
\draw[scalarnoarrow](-35:1.5) -- (-35:.5cm);
\node at (-35:1.7) {$h$};
\node (c) at (0,-0.1){};
%%inner ring
\begin{scope} [shift={(-1.5,0)}]
\begin{scope}[shift={(1.12,1.45)}, rotate=-75]
\draw[fermionloop] (2,0) arc (360:0:.5);
\end{scope}
    \begin{scope}[shift={(1.5,-0.5)}, rotate=0, scale=0.9]]
\draw[fill=black] (-0.1,-0.1) rectangle (0.1,0.1);
\end{scope}
\begin{scope}[shift={(1.92,0.27)} ,rotate=45, scale=0.9]
\draw[fill=black] (-0.1,-0.1) rectangle (0.1,0.1);
\end{scope}
\begin{scope}[shift={(1.92,-0.27)} ,rotate=45, scale=0.9]
\draw[fill=black] (-0.1,-0.1) rectangle (0.1,0.1);
\end{scope}
\begin{scope}[shift={(1.08,-0.27)} ,rotate=45, scale=0.9]
\draw[fill=black] (-0.1,-0.1) rectangle (0.1,0.1);
\end{scope}
\begin{scope}[shift={(1.08,0.27)} ,rotate=45, scale=0.9]
\draw[fill=black] (-0.1,-0.1) rectangle (0.1,0.1);
\end{scope}
\node (c) at (0,-0.1){};
\end{scope}
\end{tikzpicture} (=\hat G)
\\&+ \cdots \,.
\end{split}
\end{align}
If the Yukawa threshold correction $\Delta {y_q}$ is evaluated only at the
order $g_3^2$, this operation generates a full-model parametrized
expression including all orders $\ord( y_q^4 g_3^{2n})$.\footnote{%
 Eq.\ \eqref{eq:expand_Gamma_h4_SM} can be connected to the remark of
 footnote \ref{footnotenonpolynomial}: The 1-loop diagram
 evaluated at $v\ne0$ involves terms logarithmic in the
 quark mass, $\left.i\Gamma_{h^4}^{\eft}\right|_{\hat y_q^4} \propto
 \hat y_q^4 \log (\hat m_q /Q)$. Thus the contributions resulting
 from the double loop expansion applied to such 1-loop contribution
 contain terms of the structure
\begin{align}
	\label{eq:rep_logmQ}
	\left[e^{\Delta {y_q} \frac{\partial}{\partial \hat y_q}} \log\frac{\hat y_q v}{\sqrt{2}Q}\right]_{\hat y_q={y_q} f_q(\beta)} = \log \frac{m_q}{Q} -\sum_{i}\left(-\frac{\Delta {y_q}}{{y_q}}\right)^i,
\end{align}
where the ratio in the last term  can be simplified as $\frac{\Delta
  {y_q}}{{y_q}}\propto g_3^2 x_q+\ldots$. The arising terms explicitly
break the connection between Yukawa couplings $y_q$ and the $x_q$
parameter which exists for $\tilde \Gamma$, i.e.~the power of $x_q$ in
$\Gamma$ is not determined by the power of $y_q$ but is associated also with
higher orders in $g_3^2$.
  }
On the far r.h.s.\ of \eqref{eq:expand_Gamma_h4_SM} we singled out a particular contribution, denoted
as $\hat G$ and involving four black boxes at the vertices and one
black box on a quark line. The  black vertex boxes  indicate that the Yukawa coupling
in front of the trilinear vertex $\bar qqh$ is replaced
by the threshold correction $\Delta {y_q}$ with the $x_q$ dependence discussed
 in eq.~\eqref{eq:constraint_Delta_y}. \footnote{The effective vertices and the threshold correction may differ by contributions from the conversion of the regularization scheme ( \DRbar and \MSbar) between the full-model and the EFT.
 These contributions do not introduce any $x_q$ dependence at $\ord(g_3^2)$.}
The black box in the quark propagator represents a mass insertion $ \Delta m_q = v\Delta {y_q}/\sqrt{2} $.\footnote{Note that threshold corrections to the VEV $v$ do not contribute
at $\ord( g_3^2)$.}
Both threshold corrections
$\Delta y_q$ and $\Delta m_q$ contain a term proportional to $g_3^2
x_q$, whose coefficient
coincides with  $a_{(1,1)}$
in  eq.~\eqref{eq:1-loop-as}.

The explicit contribution of
the form $\hat G$ is evaluated as
\begin{align}
\hat G =  (-i\Delta m_q) (-i\Delta {y_q})^4 \int_{k} f(k,m_q),
\end{align}
where the integrand $f(k,m_q)$ is the same as the counterpart for $G$ in eq.~\eqref{eq:G_contribution}. 
At $\ord( g_3^2 x_q)$ the threshold corrections $(-i\Delta y_q)$ and 
$(-i\Delta m_q)$ coincide with the effective vertices $\Delta v_{\bar qqh}$ and $\Delta v_{\bar qq}$,  respectively.
Thus,  the overall contribution of $\ord( y_q^4 g_3^{10} x_q^5)$ in $G$ and in $\hat G$ is the same.
In general, all contributions to $\Gamma_{h^4}^{\eft}$ at
$\ord( y_q^4 g_3^{10} x_q^5)$ arise from diagrams similar to $\hat G$, 
with one fermion loop and five black boxes,
where at least one internal fermion propagator is dressed by a black
box. All such contributions involve the factors  $(\Delta m_q)^{n}
(\Delta y_q)^{(5-n)} $  with $1\leq n\leq5$.

There is clearly a
one-to-one correspondence between such generalized terms of type $G$
and $\hat G$, i.e.\ on the l.h.s.\ and r.h.s.\  in
eq.~\eqref{eq:matching_constraint}. Hence the  $x_q^5$ terms cancel in
this equation. After this cancellation,
extracting the threshold correction $\Delta\lambda$ from
eq.\ \eqref{eq:matching_constraint} leads to a result compatible
with the constraints of sec.\ \ref{sec:threshold}.

\section{Reparametrization of threshold corrections in the MSSM-SM matching}
\label{sec:resummation}
In this section we will illustrate the resummation
implied by the constraints on leading $x_q$ contributions presented in \secref{sec:constraint}.
As a matter of principle the threshold corrections may be expressed in parameters of the full-model  or
of the EFT.
Evaluating matching corrections in full-model parametrization, followed by reparametrization in terms of EFT parameters, leads to the announced resummation of
 leading $x_q$ contributions to observables.
 We will precisely specify the structure of terms covered by the resummation and provide several applications which go beyond
 ref.~\cite{Kwasnitza:2020wli} on the Higgs boson mass correction.

\subsection{What can be resummed}
\label{sec:what_resum}
The structure of the matching corrections result in the form $\hat y_q = f(y_q,g_i
)$ and $\hat \lambda= \bar f(y_q
,g_i
)$,
where again full-model parameters are denoted without hat, SM
parameters with hat. The functions $f$ and $\bar f$ are constrained by  eqs.~(\ref{eq:constraint_Delta_y},\ref{eq:constraint_Delta_lam}).
In the calculation of the Higgs mass, the SM 
parameter $\hat\lambda$ is predicted
while $\hat y_q$ is instead determined via experimental low-energy observables.
One therefore needs the inverted relation for the MSSM Yukawa coupling
${y_q}= \frac{1}{f_q(\beta)}\left[\hat y_q+\Delta y^{\SM}_q\right]$ and
use it to express the threshold correction for $\hat \lambda$ in terms
of SM couplings. Using the constraints on $x_q$ this leads to the
following towers of terms:\footnote{The elimination of MSSM gauge couplings $g_{1,2,3}$ in terms of EFT parameters works analogously.} 
\begin{align}\label{eq:Delta_lam_rep}
&\hat\lambda^{\SM}\equiv
	\left.\hat\lambda
	\right|_{{y_q}= \frac{1}{f_q(\beta)}\left[\hat y_q+\Delta y^{\SM}_q\right] } \\*
\label{eq:triangle_resum}
\begin{split}
&\supset
\hat A\phantom{ii} \left(\propto x_q^{\leq m}\, \right)\\[2.5ex]
&+
 \hat A \hat g_3^2 \left(\propto x_q^{\leq m}\, +\, 
\tikzmark{l1} 
 \phantom{i}k_{(m+1,1)} x_q^{m+1}\right)  \phantom{iiiii}
\tikzmark{r1} \\[4.5ex]
&+\hat A \hat g_3^4\left(\propto x_q^{\leq m}\, +
\tikzmark{ll2}
 \phantom{i}k_{(m+1,2)} x_q^{m+1}  \phantom{ii}
\tikzmark{rr2}
+ \phantom{i} k_{(m+2,2)} x_q^{m+2}\right)\\[4.5ex]
&+ \hat A
\hat g_3^6 \left(\propto x_q^{\leq m}+k_{(m+1,3)} x_q^{m+1} 
\phantom{i}+
k_{(m+2,3)} x_q^{m+2}\phantom{iii}\,
+\phantom{i}
k_{(m+3,3)} x_q^{m+3}\right)\\[1.5ex]
&+\cdots\\[1.5ex]
&+ \hat A\hat g_3^{2l}\left(
\cdots +k_{(m+l-3,l)} x_q^{m+l-3}
+k_{(m+l-2,l)} x_q^{m+l-2}+
\tikzmark{ll3}
k_{(m+l-1,l)} x_q^{m+l-1}
\phantom{ii}+
\tikzmark{rr3}
\phantom{ii}
\tikzmark{l3} 
 k_{(m+l,l)} x_q^{m+l}\right), \phantom{iii}
\tikzmark{r3} 
\\[1.5ex]
&+\cdots
\end{split}
\begin{tikzpicture}[remember picture,overlay]
\draw[blue,loosely dashed](l1) edge [out=120, in=150] (r1);
\draw[blue,loosely dashed](l1) to (l3);
\draw[blue,loosely dashed](r1) to (r3);
\draw[orange,loosely dashdotted](ll2) edge [out=120, in=140] (rr2);
\draw[orange,loosely dashdotted](ll2) to (ll3);
\draw[orange,loosely dashdotted](rr2) to (rr3);
\end{tikzpicture}
\end{align}
where $\hat A$ can be a combination of couplings such as $\{\hat y_q^4, \hat y_q^2\hat g_1^2, \hat  y_q^2\hat g_2^2, \hat y^6_q\}$
and $m$ represents the maximal power of $x_q$ terms in the 
corresponding full-model corrections $m\in\{4,2,2,6\}$, 
see constraint from eq.~\eqref{eq:constraint_Delta_lam}.
The terms in the first column in eq.~\eqref{eq:triangle_resum}  arise already in full-model
parametrization, while all other terms appear only in EFT-parametrization through reparametrization.
The terms encircled in blue are already fixed by the first term in eq.~\eqref{eq:triangle_resum} with reparametrization at leading order.
The orange terms are determined by the first two lines of  eq.~\eqref{eq:triangle_resum} together with reparametrization at next-to-leading order.

To illustrate this resummation we show examples in \tabref{tab:xf-resumm}.
In \tabref{tab:xf-resumm}, the coupling structure
$ \hat A$ and the corresponding resummation
terms out of  eq.~\eqref{eq:triangle_resum} are specified.

\begin{table}[h]
	\centering
	\begin{tabular}{cccccc}
		\toprule
		loop order & {\color{blue}$\left.\ \hat\lambda^{\SM}
			\right|_{\hat y_q^4
				 \hat g_3^{2l}}$} 
		& {\color{blue} 
		$\left.\hat \lambda^{\SM}\right|_{\hat g_{1,2}^2 \hat 
			 y_q^2  \hat g_3^{2l}}$}
		& {\color{blue}$\left.\hat \lambda^{\SM}\right|_{\hat y^6_q \hat g_3^{2l}}$}
		&	{\color{orange}	$\left.\hat \lambda^{\SM}\right|_{\hat  y_q^4 \hat g_3^{2l}}$}
		&$\cdots$
		\\
		\midrule
		\twoL & $\hat y_q^4 \hat g_3^{2} x_q^5 $ & $\hat g_{1,2}^2 \hat 
		 y_q^2  \hat g_3^{2} x_q^3 $
		& -
		& -
		&$\cdots$
		\\[0.5em]
		\thrL & $\hat y_q^4 \hat g_3^{4} x_q^6 $ & $\hat g_{1,2}^2 \hat 
		 y_q^2  \hat g_3^{4} x_q^4$
		&$\hat y_q^6  \hat g_3^{2} x_q^7 $
		& $\hat y_q^4 \hat g_3^{4} x_q^5 $
		&$\cdots$
		\\[0.5em]
		\fourL & $\hat y_q^4 \hat g_3^{6} x_q^7 $ & $\hat g_{1,2}^2 \hat 
		 y_q^2  \hat g_3^{6} x_q^5$		
		&$\hat y_q^6 \hat g_3^{4} x_q^8 $
		& $\hat y_q^4 \hat g_3^{6} x_q^6 $
		&$\cdots$
		\\
		\vdots & \vdots & \vdots 		& \vdots
		& \vdots 		&$\ddots$
		\\
		\bottomrule
	\end{tabular}
	\caption{
		Contributions to $\hat \lambda^{\SM}$ of the highest (blue) and next-to-highest (orange) power in
		the $x_q$ parameter which exist at their respective 
		order if parametrized by SM parameters. The orange terms 
		have been given only for the choice $\hat A = \hat y_q^4$.
		But corresponding structures for other terms can be given explicitly.}
	\label{tab:xf-resumm}
\end{table}

The statements on the resummation of $x_q$ can be converted into statements on the resummation of the UV parameters $A_q$ and $\tan \beta$.
In our notation the soft-parameter $A_q$ is absorbed in the quantities $X_q$ and $Y_q$, see eq.~\eqref{eq:trilinear_exp}. Therefore, the terms of highest 
power in $A_q/M_S$ 
are in one-to-one correspondence with the appearance of highest power 
contributions of $x_q$.
Now we turn to the appearance of leading powers of $\tan\beta$ in our conventions.
There are two mechanisms which introduce positive powers in the parameter $\tan \beta$ relevant for our discussion.
\begin{itemize}
		 \item Explicit  chiral squark flips in sbottom lines introduce a factor of
	$\tan \beta$ through $x_b$. 
	 \item Additional $\tan\beta$ proportionality originates from an interaction
	 between (down-type) quarks/squarks and
	 (up-type) Higgs scalars proportional to the bottom Yukawa coupling and the ``wrong'' trigonometric functions, i.e.~${\propto y_b s_\beta  \approx \hat y_b \tan\beta  }$.
\end{itemize} 
Despite the additional possible appearance one could follow
 the arguments given in \secref{sec:proof} to conclude that in full-model parametrization the
(unsuppressed) maximal power of $\tan \beta$ in threshold corrections is 
bounded by the power of the down-type Yukawa coupling $y_b$, analogous to the $x_b$ parameter.
This means that the well-known $\tan\beta$-resummation
in the bottom Yukawa threshold $\Delta y_b^\SM$
extends to $\hat\lambda^{\SM}$ at orders which contain powers of the bottom Yukawa coupling.

\subsection{Applications}
\label{sec:application}
Here we will derive all-order formulas for the resummation terms such
as the ones in eq.~\eqref{eq:triangle_resum}.

\paragraph{\underline{$\hat\lambda^\SM$ at $\{\hat y_q^4 \hat g_3^{2l} x_q^{4+l}, \hat
		y_q^2\hat g_1^2 \hat g_3^{2l} x_q^{2+l}, \hat  y_q^2\hat g_2^2\hat g_3^{2l} x_q^{2+l}\}$}:}
We begin with the resummation of these terms, which correspond to the second and third column of \tabref{tab:xf-resumm},
 corresponding to the blue terms
in eq.~\eqref{eq:triangle_resum} with $\hat A\in\{\hat y_q^4, \hat
y_q^2\hat g_1^2, \hat  y_q^2\hat g_2^2\}$.
Ref.~\cite{Kwasnitza:2020wli}
has used this resummation and presented 2-loop and 3-loop terms. Here we give predictions generalized to all orders in QCD contributions.

In order to achieve resummation of these terms it is 
sufficient to consider the highest-power $x_q$ terms in 
specific 1-loop threshold contributions allowed by the 
constraints in
 eqs.~(\ref{eq:constraint_Delta_y},\ref{eq:constraint_Delta_lam}) 
which may be written more explicitly as
\begin{subequations}
\label{eq:thresh_1l}
\begin{align}
	\Delta {y_q} &\supset y_q f(\beta) g_3^2 a_{(1,1)}x_q~,  \\
	\Delta \lambda &\supset (y_q f_q(\beta))^4 c_{(4,0,{y_q})} x_q^4 + 
	 (y_q f_q(\beta))^2  g_{1}^2 c_{(2,0,1)} x_q^2 +
	 (y_q f_q(\beta))^2  g_{2}^2 c_{(2,0,2)} x_q^2~,
\end{align}
\end{subequations}
where $a_{(1,1)}$ is the 1-loop coefficient of the highest possible
$x_q$ term in the threshold correction $\Delta {y_q}$. Similarly,
 the 1-loop constants $c_{(m,0,\cdots)}$ 
represent the coefficients of the highest powers in $x_q$,
i.e.~$m\in\{4,2,2\}$, appearing at their respective 
order $\{\ord(y_q^4), \ord( y_q^2  g_1^2), \ord( y_q^2  g_2^2)\}$. 
We note that $c_{(2,0,1)}$ and $c_{(2,0,2)}$
depend on the angle $\beta$ due to the $D$-term interaction of squarks
with the SM-like Higgs boson $\tilde q^2h^2$. In
contrast $c_{(4,0,{y_q})}$ and $a_{(1,1)}$ are independent  of
$\beta$.

Combining the threshold corrections as indicated in eq.~\eqref{eq:Delta_lam_rep},
that is eliminating  the MSSM coupling $y_q$ by the SM parameter $\hat y_q$, results in
\begin{align}
\label{eq:leading-loop_coeff_yt}
\left.\hat\lambda^{\SM}\right|_{\hat y_q^4 \hat g_3^{2l} x_q^{4+l}}&
=\hat y_q^4 x_q^4
\frac{c_{(4,0,y_q)}}{\left[1 +  \hat g_3^{2}   a_{(1,1)} x_q\right]^4}  ,\\
\label{eq:leading-loop_coeff_g12}
\left.\hat\lambda^{\SM}\right|_{\hat g_{1,2}^2 \hat 
	 y_q^2  \hat g_3^{2l} x_q^{2+l}}&
=\hat y_q^2 x_q^2
\frac{\hat g_{1}^2 c_{(2,0,1)} + \hat g_2^2 c_{(2,0,2)} }{\left[1 +  \hat g_3^{2}   a_{(1,1)} x_q\right]^2}  .
\end{align}
which is the explicit formula of the resummed terms of this kind. 

By considering the top sector, the discussed procedure allows for 
a resummation of the highest-power $x_t=X_t/M_S$ terms. 
This fact was first pointed out in  ref.~\cite{Kwasnitza:2020wli}, and
ref.~\cite{Kwasnitza:2020wli} predicted the 2-loop and 3-loop term at
$\ord(\hat y_t^2\hat g_{1,2}^2 \hat g_3^2 x_t^3, \hat y_t^2\hat
g_{1,2}^2 \hat g_3^4 x_t^4)$, orders which mix electroweak and QCD corrections.
During the preparation of ref.~\cite{Kwasnitza:2020wli} the full threshold correction at $\ord(\hat y_t^2\hat g_{1,2}^2 \hat g_3^2)$ was calculated in ref.~\cite{Bagnaschi:2019esc}.
Both results agree exactly.
Furthermore, the 2-loop contribution at $\ord(\hat y_t^4\hat g_3^2 x_t^5)$ was calculated first in ref.~\cite{Bagnaschi:2014rsa}
and is correctly reproduced by eq.~\eqref{eq:leading-loop_coeff_yt}, thereby confirming our analyses.

For the bottom corrections in eq.~\eqref{eq:leading-loop_coeff_yt}, this means that the highest powers of the fundamental parameters $A_b/M_S $ and $ \mu\tan \beta /M_S$
in $\hat\lambda^{\SM}$ are resummed, i.e. \break
 ${\ord (\hat y_b^4 \hat g_3^{2l} \left( A_b/M_S\right)^{l_1}\left(\mu\tan \beta/M_S\right)^{l_2})}$ with $l_1 +l_2=l+4$.

\paragraph{\underline{$\hat\lambda^\SM$ at $\hat y_q^4 \hat g_3^{2l} x_q^{3+l}$:}}
Now we focus on the terms listed in the fifth
column of \tabref{tab:xf-resumm}. More specifically,
we consider further contributions, which are independent of $g_{1,2}$
but otherwise contain \eqref{eq:thresh_1l} plus new terms \footnote{In principle a 1-loop term in $\Delta \lambda\propto y_q^4 x_q^3$ would be relevant for the discussion. However, in a direct calculation such term is absent at $\ord((v/M_S)^0)$.}
 \begin{subequations}
\label{eq:thresh_2l}
\begin{align}
\Delta \lambda &\supset  (y_q f_q(\beta))^4  c_{(4,0,{y_q})} x_q^4  + (y_q f_q(\beta))^4 g_3^2 c_{(4,1,{y_q})} x_q^4~,\\*
\Delta {y_q} &\supset y_q f_q(\beta) g_3^2\left( a_{(0,1)}+ a_{(1,1)}x_q \right)+  y_q f_q(\beta) g_3^4 a_{(1,2)}x_q~,\\*
\Delta g_3 &\supset g_3^3 \delta_{g_3}~,
\end{align}
\end{subequations}
where the new coefficient $ a_{(0,1)}$
corresponds to a 1-loop term subleading in $x_q$, and the new coefficients
$c_{(4,1,{y_q})}$ and $a_{(1,2)}$ to leading 2-loop terms.
The term $g_3^3 \delta_{g_3}$ denotes the 1-loop threshold correction for the QCD coupling.
Solving the 2-loop matching equations for the SM couplings $\hat g_3$
and $\hat y_q$ without truncation resums the orange circled terms of
eq.~\eqref{eq:triangle_resum}.

Consequently, taking into account the terms of
eq.~\eqref{eq:thresh_2l} in full-model parametrization allows to resum
contributions which are leading and subleading in powers of $x_q$ into $ \hat\lambda^{\SM}$.
Analogously to eq.~\eqref{eq:leading-loop_coeff_yt}, we present a closed version for the  subleading $x_q$ contributions in
terms of the coefficients $c_{(\cdots)}$, $a_{(\cdots)}$ and $\delta_{g_3}$.
\begin{align}
\label{eq:subleading-loop_coeff_yt}
\begin{split}
\left.\hat\lambda^{\SM}\right|_{\hat y_q^4 \hat g_3^{2l} x_q^{3+l}}
=\hat y_q^4 \hat g_3^{2} x_q^4
\frac{c_{(4,1,y_q)}\left(1 +  \hat g_3^{2}   a_{(1,1)} x_q \right)-4c_{(4,0,y_q)}\left(a_{(0,1)} +  \hat g_3^{2} \left(a_{(1,2)} -2 \delta_{g_3}  a_{(1,1)} \right) x_q\right)}{\left[1 +  \hat g_3^{2} a_{(1,1)}x_q \right]^5}.
 \end{split}
\end{align}

In the case of top-Yukawa contributions, the inclusion of the second and third 
column of \tabref{tab:xf-resumm} was discussed and implemented in the version of the code \feft of ref.~\cite{Kwasnitza:2020wli}.
Actually, the code implemented all corrections in  eq.~\eqref{eq:thresh_2l} in full-model parametrization.
Hence, we can identify here that the code of ref.~\cite{Kwasnitza:2020wli} automatically
resums  all terms in
eq.~\eqref{eq:subleading-loop_coeff_yt}
and thus includes all terms in the fifth column of \tabref{tab:xf-resumm}.

By the inclusion of the terms in the second, third and fifth
column of \tabref{tab:xf-resumm}, ref.~\cite{Kwasnitza:2020wli} illustrated that for
high values of the  parameter $X_t/M_S$
that numerical convergence
of perturbation series is improved.

\paragraph{\underline{$\hat\lambda^\SM$ at $\hat y_t^6 \hat g_3^{2l} x_t^{6+l}$:}}
These terms have more factors
of the Yukawa couplings and 
their resummation corresponds to the fourth
column of \tabref{tab:xf-resumm}. 
We focus on the top sector and more specifically on the resummation of the stop-mixing parameters
$x_t= X_t/M_S$.
In order to resum such contributions,
we consider the following loop corrections in 
the high scale matching procedure \footnote{Note that there exists no term $y_t^3 x_t^3$ in the 1-loop Yukawa threshold correction.}
\begin{subequations}
\label{eq:thresh_2l_yuk}
\begin{align}
\Delta \lambda &\supset (y_t s_\beta)^4 c_{(4,0,y_t)} x_t^4 +(y_t s_\beta)^6
 c_{(6,0,y_{t2})} x_t^6\,,\\
\Delta y_t &\supset y_t s_\beta g_3^2 a_{(1,1)}x_t  + (y_t s_\beta)^3 a_{(2,0,y_t)}x_t^2  
+  (y_t s_\beta)^3 g_3^2  a_{(3,1,y_t)}x_t^3\,,
\end{align}
\end{subequations}
where we substituted the coupling $Y_t \cot\beta$  by the  parameters $X_t$ and $\mu$ which introduces a dependence on  $\cot\beta$ in the new coefficients of eq.~\eqref{eq:thresh_2l_yuk}.
Analogous to the previous paragraphs, we invert the matching relation of the Yukawa coupling.
Neglecting terms of $\ord(\hat y_t^5)$, the expansion of the cubic equation leads to
\begin{align}
\begin{split}
y_t=&
\hat y_t\frac{1}{1 +  \hat g_3^{2}   a_{(1,1)} x_t}
-	 \hat y_t^3 \frac{a_{(2,0,y_t)}x_t^2+  \hat g_3^{2}  a_{(3,1,y_t)} x_t^3}
{\left[1 +  \hat g_3^{2}  a_{(1,1)}  x_t\right]^4}+ \cdots ~,
\end{split}
\end{align}
where the dots denote loop contributions with lower powers of $x_t$. By inserting the relation in \eqref{eq:Delta_lam_rep}, one obtains the resummed contributions as
\begin{align}
\left.\hat\lambda^{\SM}\right|_{\hat y_t^6 \hat g_3^{2l} x_t^{6+l}}&
=\hat y_t^6 \hat x_t^6
\frac{c_{(6,0,y_{t2})} \left(1+\hat g_3^{2} a_{(1,1)}  x_t \right)-4c_{(4,0,y_t)}
\left(a_{(2,0,y_t)} + \hat g_3^{2} a_{(3,1,y_t) } x_t \right)}{\left[1 +  \hat g_3^{2}   a_{(1,1)} x_t\right]^7} .
\label{eq:subleading_res}
\end{align}
Now we discuss the 3-loop term of $\ord(\hat y_t^6 \hat g_3^{2} x_t^{7})$  with the highest $x_t$ power in the degenerate mass case. The 1-loop coefficients
can be found for example in ref.~\cite{Draper:2013oza}; in the 
limit where all \SUSY masses are equal they read $c_{(4,0,y_t)} =-\kappa/2$, $a_{(1,1)} =-\kappa 4/3$ and $a_{(2,0,y_t)} =-\kappa/4$ with $\kappa =1/(4\pi)^2$. At 2-loop, $c_{(6,0,y_{t2})}$ was listed in ref.~\cite{Vega:2015fna,Kwasnitza:2020wli} and $a_{(3,1,y_t)}$ was presented in 
ref.~\cite{Mihaila:2016lfc}.
At large values for  $\tan \beta$,
the 2-loop coefficients and the resulting 3-loop correction of interest reduce to
\begin{align}
\frac{c_{(6,0,y_{t2})}}{\kappa^2}
=& -  \left[1 + \frac{1}{4}\left(19+96K \right)\cot^2\beta\right]
\approx - 1 +\ord\left(\cot \beta\right), 
\\
\label{eq:2-loop_yt}
\frac{a_{(3,1,y_t)}}{\kappa^2}
=& -  \frac{1}{9}\Big[3 \left(1+4L_S \right) +4\left(4+3 L_S-18 S_2\right) \cot^2\beta\Big]
\approx - \frac{1}{3} +\ord\left(\cot \beta, L_S\right), 
\\
\label{eq:subleading_resuma}
	\begin{split}
\left.\hat\lambda^{\SM}\right|_{\hat y_t^6 \hat g_3^{2} x_t^7}= &-\kappa^3 \hat y_t^6 \hat g_3^2 x_t^7\frac{2}{9}\Big[12\left(5+ L_S\right) +\left(187+864K-72 S_2 + 12 L_S \right)\cot^2\beta\Big]
	\\
	=&-\kappa^3\hat y_t^6 \hat g_3^2 x_t^7\frac{40}{3} +\ord\left(\cot \beta, L_S\right),
\end{split}
\end{align}
with $L_S= \log (Q^2/M_S^2)$, $S_2 =0.260434$ and $K=-0.1953256$.

The Higgs-mass calculation \feft presented in ref.~\cite{Kwasnitza:2020wli} implemented all corrections from
eq.~\eqref{eq:thresh_2l_yuk} except for the $a_{(3,1,y_t)}$ term,
which is the leading power $x_t$ term to $\Delta y_t$ at the 2-loop
order $\ord(y_t^3g_3^2)$. 
Therefore the resummation of eq.~\eqref{eq:subleading_res} is incomplete.
To inspect the numerical impact we modified the high-scale 
matching relation of the \feft code by  3-loop contributions in two versions. 
In the first version the matching is extended by the $a_{(3,1,y_t)}$ term, such that this version fully
contains the correct term presented in
eq.~\eqref{eq:subleading_resuma}.
The second modified code is constructed such that the threshold correction $\Delta \lambda$ vanishes at $\ord(\hat y_t^6\hat g_3^2x_t^7)$ exactly.
\begin{figure}
	\centering
	\includegraphics[width=0.59\textwidth]{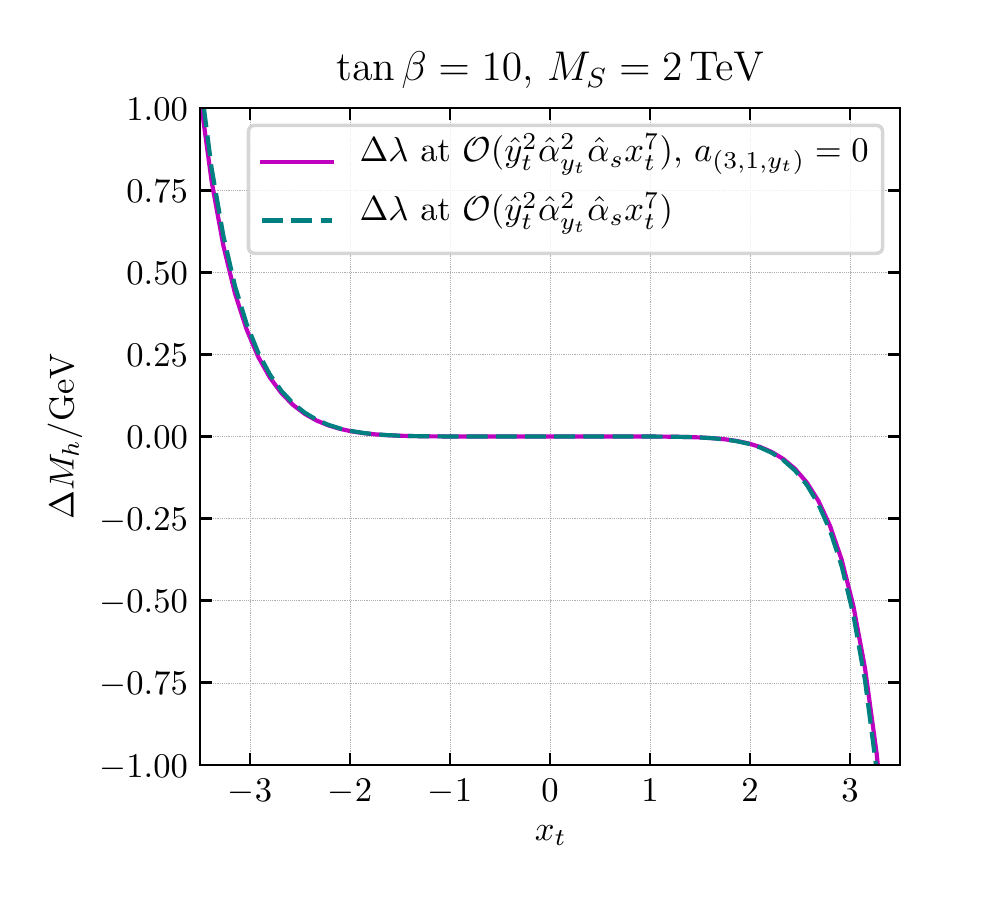}
	\caption{Impact on the Higgs mass from 3-loop threshold correction
		$\Delta\lambda$ from the highest power  $x_t$ contribution at $\ord(\hat y_t^6\hat g_3^{2})$.
	The values of the Higgs mass are obtained by the calculation implemented in ref.~\cite{Kwasnitza:2020wli} and its modifications.}
	\label{fig:xtscanrepara5tev3l}
\end{figure}
In \figref{fig:xtscanrepara5tev3l} we compare the Higgs mass as predicted by the unmodified \feft calculation (with $a_{(3,1,y_t)}=0$) and the version with the 
correct term from eq.~\eqref{eq:subleading_resuma} for a
characteristic \SUSY scale $M_S =2\,$TeV.
From both calculations we subtracted the result obtained
by the version where the high-scale correction $\ord(\hat y_t^6\hat g_3^2x_t^7)$ is set to zero.
The dashed line thus shows the
impact of the complete threshold correction eq.~\eqref{eq:subleading_resuma} on the Higgs mass.
The solid line represents the impact of the incomplete threshold correction in the implementation
of  ref.~\cite{Kwasnitza:2020wli}, where the 2-loop coefficient $a_{(3,1,y_t)}=0$.
In the considered limit the threshold correction is important and
$\Delta M_h \approx 500\,$MeV for $x_t\approx3$. 
But the difference between the complete and incomplete version, i.e.\ of the contribution of $a_{(3,1,y_t)}$ to the threshold correction in $\hat\lambda^\SM$ is  
negligible.

Hence, although
the Higgs mass calculation implemented in ref.~\cite{Kwasnitza:2020wli} does not contain a 2-loop threshold to the Yukawa matching at $\ord(y_t^3 g_3^2)$, it 
therefore reproduces the
 3-loop term $\hat y_t^6 \hat g_3^{2} x_t^7\subset \hat \lambda^{\SM}$ to a very good numerical precision.

\paragraph{\underline{$\hat\lambda^\SM$ at $\hat y_b^4 \hat y_t^{2l} (A_t/M_S)^{l} (\tan\beta)^{4+l}$:}}

Now we discuss these mixed Yukawa expressions which are beyond the terms listed in \tabref{tab:xf-resumm}, as they do not contain any QCD coupling $g_3$.
This analysis is in line with the resummation of {$\ord (\hat y_b (\hat y_t^2 A_t \tan \beta /M_S)^l)$} in the
MSSM bottom  Yukawa coupling presented in ref.~\cite{Noth:2008tw,Noth:2010jy}.
Here we derive a simultaneous resummation of these terms in $\Delta y_b^\SM$ and $\hat\lambda^\SM$.
It is sufficient in our analysis to include the following 1-loop thresholds corrections 
\begin{subequations}
	\label{eq:thresh_pure_yuk}
	\begin{align}
	\Delta \lambda &\supset (y_bc_\beta)^4 c_{(4,0,y_b)} x_b^4, \\
	\Delta y_b &\supset (y_b c_\beta) y_t^2 a_{(b,t)}  x_t \tan\beta ,
	\end{align}
\end{subequations}
where  $a_{(b,t)}$ and $c_{(4,0,y_b)}$ are independent of $x_q$ and $\tan\beta$.
Following the reparametrization prescription of
eq.~\eqref{eq:Delta_lam_rep}, 
the  included corrections to $ \hat\lambda^\SM$ are 
\begin{align}
\label{eq:resum_tanb_yuk}
\left. \hat\lambda^\SM\right|_{\hat y_b^4 \hat y_t^{2l}x_t^l x_b^4\tan^l\beta }=\hat y_b^4 \frac{c_{(4,0,y_b)}}{\left[1+\frac{\hat y_t^2}{s_\beta^2} a_{(b,t)} x_t \tan\beta \right]^4}x_b^4\,.
\end{align}
Expressed in terms of $A_t$ and $\tan\beta$, the terms in
\eqref{eq:resum_tanb_yuk} are of
$\ord( \hat y_b^4 \hat y_t^{2l} (A_t/M_S)^l (\tan\beta)^{4+l} )$.
In consequence, eq.~\eqref{eq:resum_tanb_yuk} resums terms leading in powers of $(A_t \tan\beta /M_S)$.

Now we argue why no explicit multi-loop contributions
to eq.~\eqref{eq:thresh_pure_yuk} exist  which may affect the terms in eq.~\eqref{eq:resum_tanb_yuk}.
As discussed in the items of \secref{sec:what_resum},
the only mechanism to introduce contributions $\propto \tan \beta$
at $\ord(y_b^4 y_t^{2k})$ is through the sbottom-mixing parameter $x_b$.
In order to give rise to corrections to
eq.~\eqref{eq:resum_tanb_yuk}, the diagrams would be of $\ord ( y_b^4 x_b^{k+4} y_t^{2k} x_t^k)$, for $k>1$.
But as stated in \secref{sec:thresh_comments}, the analysis in \secref{sec:constraint} forbids such (unsuppressed) terms explicitly, because the $x_b$ parameter cannot appear with higher powers than $y_b$.

We note in passing, that the resummation covers terms as $x_b^{4+l} x_t^l$ but not $x_b^4 x_t^{2l}$, where the same overall power in $x_q$ is  distributed differently among $x_{t,b}$.
Clearly, for $\tan \beta \gg x_t$ the resummed terms in eq.~\eqref{eq:resum_tanb_yuk}, at $\ord ( \hat y_b^4  \hat y_t^{2l})$ are dominant.

\paragraph{Resummation of flavor violating stop-scharm-Higgs soft parameter in $\hat\lambda^\SM$}
As announced in the comments of \secref{sec:thresh_comments} there exist constraints for flavor-violating trilinear parameters.
Now we use such a constraint to derive a resummation of
 stop-scharm-Higgs couplings, which are not listed in
 \tabref{tab:xf-resumm} as they are not associated to the third
 generation only.
In the style of eq.~\eqref{eq:leading-loop_coeff_yt}, we present an all-order equation of enhanced flavor violating contributions from soft-breaking $A$-terms.

The impact of mixing effects between the second and third generation of squarks onto the Higgs mass has been studied for example in refs.~\cite{Heinemeyer:2004by, Heinemeyer:2004gx, AranaCatania:2011ak, Arana-Catania:2014ooa} at 1-loop and in ref.~\cite{Goodsell:2015yca} at 2-loop.
It was shown that among all sources of
flavor violation the trilinear chirality 
changing interactions can particularly 
contribute to the Higgs mass 
without being in conflict with other observables.

In this section we work in the weak basis
for quarks and squarks and with mass eigenstates of the scalar Higgs sector.
The interaction of interest is induced by the 
(real) soft \SUSY-breaking trilinear coupling
\begin{align}
	\Lagr^{\text{soft}} \supset -\frac{T_{23}^u s_\beta }{\sqrt{2}}\,h \,
	\tilde c_L^\dagger 	\tilde t_R^{\phantom{\dagger}}
	-\frac{T_{32}^u s_\beta }{\sqrt{2}}\,h \,
	\tilde t_L^\dagger 	\tilde c_R^{\phantom{\dagger}}
	+ -\frac{T_{33}^u s_\beta }{\sqrt{2}}\,h \, \tilde t_L^\dagger 	\tilde t_R^{\phantom{\dagger}}
	+ \text{h.c.} ~,
\end{align}
where we decompose the $T^u_{ij}$ coupling in a product of the ($i$,$j$) entries of the up-type Yukawa ($3\times 3$)-matrix $y^u_{ij}$ and of the  $M_S$-enhanced matrix $A^u_{ij}$, i.e.~$ T_{ij}^u=y_{ij}^u A_{ij}^u$. 

In the following we elaborate how a 1-loop analysis allows
for the inclusion of contributions which are of highest power in
$(A_{23}^u A_{33}^u/M_S^2)$ in leading QCD order.
We consider matching of Green functions $\Gamma_{\bar t t h}$, $\Gamma_{\bar t c h}$ and $\Gamma_{h^4}$, where the external quarks $c$ and $t$ are the up-type quarks of the second and third generation defined in the weak basis.
Analogous to eq.~\eqref{eq:yuk_match} the matching results
in a relation of the Yukawa couplings in the SM and MSSM which is expanded in loops,  
$\hat y^u_{ij}=  s_\beta y^u_{ij} + \Delta y^u_{ij}$.
As in eq.~\eqref{eq:thresh_1l},  1-loop threshold corrections for
the Yukawa couplings $\hat y_{33}^u$, $\hat y_{23}^u$ and the quartic $\hat \lambda$ include enhancements by the trilinear parameters $A_{ij}^u$.
We focus on 1-loop contributions which contain $A_{33}^u/M_S$ and $A_{23}^u/M_S$
\begin{align}
\label{eq:higgs_T33}
\left.\Delta y^u_{33} \right|_{A^u_{33}}
&= s_\beta  y_{33}^u g_3^2 a_{(\tilde t, \tilde t)} \frac{A_{33}^u}{M_S}\supset
\begin{tikzpicture}[baseline=(c)]
\begin{scope} [shift={(0,0)}]	
\node (c) at (0,-0.1){};
\node at (-.7,.) {$ t$};	
\node at (.,.8) {$\tilde t_R$};
\node at (2.2,.8) {$\tilde t_L$};
\node at (2.9,.) {$t$};
\draw[fermion] (-0.5,0)--(.2,0);
\draw[fermionnoarrow] (.1,0)--(2.7,0);
\draw[fermion] (2.3,0)--(2.7,0);
\draw[scalarnoarrow] (.1,0) arc (180:90:1);
\draw[scalarnoarrow] (1.1,1) arc (90:0:1);
\node at (1.2,-0.3) {$\tilde g$};
\node at (1.4,1.5) {$h$};
\draw[scalarnoarrow] (1.1,1)--(1.1,1.7);
\end{scope}
\end{tikzpicture}
+\cdots\, ,
\\
\label{eq:higgs_T32}
 \left.\Delta y^u_{23} \right|_{A^u_{23}}
 &= s_\beta  y_{23}^u g_3^2 a_{(\tilde t, \tilde c)} \frac{A_{23}^u}{M_S}\supset
\begin{tikzpicture}[baseline=(c)]
\begin{scope} [shift={(0,0)}]	
\node (c) at (0,-0.1){};
\node at (-.7,.) {$ t$};	
\node at (.,.8) {$\tilde t_R$};
\node at (2.2,.8) {$\tilde c_L$};
\node at (2.9,.) {$c$};
\draw[fermion] (-0.5,0)--(.2,0);
\draw[fermionnoarrow] (.1,0)--(2.7,0);
\draw[fermion] (2.3,0)--(2.7,0);
\draw[scalarnoarrow] (.1,0) arc (180:90:1);
\draw[scalarnoarrow] (1.1,1) arc (90:0:1);
\node at (1.2,-0.3) {$\tilde g$};
\node at (1.4,1.5) {$h$};
\draw[scalarnoarrow] (1.1,1)--(1.1,1.7);
\end{scope}
\end{tikzpicture}
+\cdots\, ,
\\
\label{eq:higgs_T32_quartic}
\left.\Delta \lambda\right|_{(A_{33}^u A^u_{23})^2}&=
 s_\beta^4 (y^u_{33} \,y^u_{23})^2c_{(\tilde t, \tilde c)}  \left( \frac{A^u_{33} A^u_{23}}{M_S^2} \right)^2   \supset
\begin{tikzpicture}[baseline=(c)]
\begin{scope} [shift={(-1.5,-0.7)}]
		\draw[scalarnoarrow] (-.8,2.) -- (0,1.5);
\draw[scalarnoarrow] (0,1.5) -- (1.5,1.5);
\draw[scalarnoarrow] (1.5,1.5) -- (2.3,2.);
\node at (-1,2.) {$h$};
\node at (2.5,2.) {$h$};
\node at (.75,1.8) {$\tilde t_R$};
\node at (1.85,.75) {$\tilde t_L$};
\draw[scalarnoarrow] (-.8,-.5) -- (0,0);
\draw[scalarnoarrow] (0,0) -- (1.5,0);
\draw[scalarnoarrow] (1.5,0) -- (2.3,-.5);
\node at (-1,-.5) {$h$};
\node at (2.5,-.5) {$h$};
\node at (.75,-.3) {$\tilde t_R$};
\node at (-.3,.75) {$\tilde c_L$};
\draw[scalarnoarrow] (0,1.5) -- (0,0);
\draw[scalarnoarrow] (1.5,1.5) -- (1.5,0);
\end{scope} 
\node (c) at (0,-0.1){};
\end{tikzpicture}+\cdots,
\end{align}
where the coefficients $a_{(...)}$ and  $c_{(...)}$ are independent of $A_{ij}^u$ and $\beta$.
Solving the matching equation for the MSSM coupling $y_{ij}^u$, including 1-loop corrections from
eqs.~(\ref{eq:higgs_T33},\ref{eq:higgs_T32}), results in a power series in $(g_3^{2n}( A_{33}^u,A_{23}^u)^n)$.
In consequence, the expansion of $\Delta\lambda$ in eq.~\eqref{eq:higgs_T32_quartic} in terms of
SM couplings leads to a series of (unsuppressed) highest power contributions in $A_{33}^u$ and $A_{23}^u$
\begin{align}
\label{eq:charmstopresum}
	\left.\hat\lambda^{\SM}\right|_{(\hat y_{33}^u \hat y_{23}^u)^2 \hat g_3^{2l}}& = (\hat y_{33}^u \,\hat y_{23}^u)^2  \left(\frac{A_{33}^u A_{23}^u}{M_S^2}\right)^2 \frac{ c_{(\tilde t, \tilde c)}}{\left[1 +  \hat g_3^{2}  a_{(\tilde t,\tilde t)} \frac{A_{33}^u}{M_S}\right]^2
	\left[1 +  \hat g_3^{2}   a_{(\tilde t, \tilde c)} \frac{A_{23}^u}{M_S}\right]^2} .
\end{align}

Now we ask if there exist explicit
multi-loop corrections to $\Delta \lambda$ 
or to $\Delta y_{ij}^u$ (in full-model 
parametrization)  which modify terms in 
eq.~\eqref{eq:charmstopresum} 
(EFT-parametrization).
Indeed, as in all previous cases we can apply the arguments given in \secref{sec:proof}
and conclude that the power of $A_{ij}$ is maximally the power of the Yukawa coupling $y_{ij}$ in the threshold correction:

 \textit{In full-model parametrization, the highest power contributions to threshold corrections $\Delta \lambda$ 
	and $\Delta y_{ij}^u$ are of order $\ord(g_3^{2n} (y_{33}^u  A_{33}^u/M_S  )^{n_{t}} (y_{23}^u A_{23}^u/M_S )^{n_{tc}})$, for all positive integers {$n$, $n_t,n_{tc}>0$}}.
\newline This means that
 terms in eq.~\eqref{eq:charmstopresum}  do not receive any corrections from higher orders and are resummed.
 
Similar conclusions apply for contributions
to $\hat\lambda^\SM$
with no $\hat y^u_{33}$ enhancement and more powers in $\hat y_{23}^u$, i.e the inclusion of 1-loop contributions $\ord((y_{23}^u A_{23}^u)^4)$ to $\Delta \lambda$ results in a resummation of
$\ord ((\hat y_{23}^u)^4\hat g_3^{2l} (A_{23}^u/M_S)^{l+4})$ in $\hat\lambda^\SM$. Clearly, the discussion of the $A_{32}^u$ coupling is identical.

\subsection{What cannot be resummed}
So far we have mainly focused on the resummation of QCD-enhanced terms
$\hat g_3^{2n} x_q^n$. Now we discuss, whether it is possible to resum
terms $\propto\hat y_q^{2n} x_q^{2n}$, where the powers of $x_q$ are
related to the powers of Yukawa couplings.
Such terms are of particular interest, since at a given loop order they are of higher power in $x_q$ than the $g_3$-enhanced terms.
The answer is no.
As derived in \secref{sec:proof}, the maximal power of $x_q$ (unsuppressed) in explicit contributions to threshold corrections is
given by the power of Yukawa couplings $y_q$.
Thus, the reparametrization of threshold corrections in terms of EFT couplings alone cannot completely capture highest-power terms of $\ord((\hat y_t x_t)^{n_t}(\hat y_b x_b)^{n_b})$.
If no other restrictions are invoked, our analysis has implications on the limits of reorganizing the
perturbative expansion:
\begin{itemize}
	\item one cannot resum $\ord((\hat y_t x_t)^{n_t}(\hat y_b x_b)^{n_b})$ in corrections $\Delta y_q^\SM$,
	\item one cannot resum $\ord((\hat y_t x_t)^{n_t}(\hat y_b x_b)^{n_b})$ in corrections $\hat\lambda^\SM$.
\end{itemize}

Note that the discussion around eq.~\eqref{eq:resum_tanb_yuk} leads to a resummation for pure Yukawa orders unrelated to QCD,
which is consistent with the given remarks.

\section{Conclusions}

In this paper we established all-order statements on the appearance of
the parameters $x_q$ in MSSM Green functions and threshold corrections
between MSSM and SM couplings in the context of minimal subtraction schemes \DRbar and \MSbar. We focused particularly on the quartic
Higgs coupling $\lambda$. The optimum setting for the statements is
full-model parametrization, i.e.~perturbative expansion in terms of
MSSM couplings including, if needed, truncation of the expansion at
fixed order in terms of these couplings.

In full-model parametrization our first main statements are constraints on
the threshold corrections $\Delta y_q$ and $\Delta \lambda$, see
eqs.~\eqref{eq:constraint_Delta_y} and 
\eqref{eq:constraint_Delta_lam}, stating which powers in the $x_q$
parameters are forbidden. These statements generalize results from the literature 
focusing on constraints for powers of $\tan\beta$ in $\Delta y_b$
\cite{Carena:1999py, Noth:2008}. 
Our second set of statements are constraints on powers of $x_q$  in the
Green functions $\Gamma_{\bar q q h}$ and $\Gamma_{h^4}$, see
eqs.~\eqref{eq:constr_gamma_hqq} and \eqref{eq:constr_gamma_hhhh}.
These constraints are weaker than the ones on threshold corrections;
the higher-power $x_q$ contributions to the Green functions
originate from an integration region where the internal loop momenta are soft.
 The consistency of the constraints on $\Delta \lambda$ and
 $\Gamma_{h^4}$ has been  illustrated by an explicit matching calculation
 of $\Gamma_{h^4}$ in the broken phase in
 \secref{sec:expl_cancel}. There the cancellation of powers $x_q^{\geq
   4}$ contributions at $\ord(y_q^4g_3^{10})$ was explicitly  demonstrated.

 We remark that the constraints presented here are specific to
 minimal subtraction schemes. For example, the appearance of
 $x_q^n$-terms in Green-functions renormalized in the on-shell scheme
 is different, see e.g.\ \cite{Noth:2008}. 
 
 One practical relevance of the threshold constraints is that they often
 lead to an effective resummation of (sub-)leading
 $x_q$-contributions: Many  calculations are done in a context where the Yukawa couplings are fixed by
 low-scale parameters. This requires the
 full-model Yukawa coupling $ y_q$ to be determined in terms of the
 EFT coupling $\hat y_q$.
 This procedure inverts and combines fixed-order relations and thereby
 generates terms of higher orders in loops and in $x_q$. Using the
 derived constraints one can prove that certain towers of terms are
 actually correct, i.e.\ ``resummed'' at all orders.
 In \secref{sec:what_resum}  we gave a general description of the
 towers of the resummed terms in $x_q$,  or in the context of related parameters such as $A_q$ and $\tan\beta$.
 As a special case we identified that the well-known $\tan \beta$-resummation 
 in the bottom Yukawa coupling works analogously for $\hat \lambda$ and therefore for the Higgs mass.
 In \secref{sec:application} 
 we explored a plethora of multi-loop
 structures which are subject to this parameter resummation. 
 We gave analytic expressions for all-order  contributions to $\hat\lambda$ for the dominant parameter-enhanced terms,
 i.e.~(sub)leading powers of $\tan\beta,~A_q$ or $x_q$.

Here we summarize the relevant properties why this resummation of highest power terms in the BSM parameter $x_q$ is possible.
 \begin{itemize}
 	
 	\item \textit{Matching in full-model parametrization:} The
          matching of Green functions at a fixed order yields
          threshold corrections 
          where each factor of $x_q$ is necessarily accommodated by a
          factor of full-model Yukawa coupling $y_q$. 
 	\item \textit{Decoupling of $y_q$ and $x_q$ through
          reparametrization:}
          The threshold corrections are then reparametrized in terms of
          EFT parameters. Because of the structure of the
          Yukawa matching, 
        reparametrizing  decouples the maximal power of $x_q$ from the 
 	power in the EFT Yukawa coupling $\hat y_q$. 
 	As a result  the threshold corrections
 	     contain terms of $\mathcal{O}(\hat y_q(\hat g_3^2x_q)^n)$, i.e.\ terms where each additional order 
              in $g_3^2$ is accompanied by a factor of $x_q$.
 		\item \textit{Correctness of the  reparametrization terms with higher power in $x_q$:}
  By following the arguments given in \secref{sec:proof} (or in
  \secref{proof}), one can check that such terms, which are leading or
  subleading in $x_q$, are correctly ``resummed'' for any $n$, even
  though they are generated from a fixed-order calculation via
  reparametrization. 
 \end{itemize}
 	Clearly, similar analyses
may be carried out for other BSM parameters if similar properties apply.

\section{Acknowledgments}

We are grateful to Alexander Voigt for discussions on applications of our results, and to Ulrich Nierste 
for detailed discussions of 
$\tan\beta$-resummation and of refs.~\cite{Carena:1999py,Hofer:2009xb}.
This research was supported by the German Research Foundation (DFG)
under grant number STO 876/2-2 and by the high-performance computing
cluster Taurus at ZIH, TU Dresden.

\clearpage

\appendix

%Subsequent command helps against bug of overlapping text.
%\renewcommand*{\thesection}{\Alph{section}}
\section{Leading contributions from LME and effective vertices}
\label{app:1}
In this appendix we list the effective vertices appearing in a large
mass expansion (LME)
of diagrams with at least one heavy internal propagator.
First, we analyze the relevant Higgs-fermion interactions which lead to non-trivial $x_q$ contributions. Second, we discuss gluon interactions by invoking 
arguments from $SU(3)_C$ BRST invariance.
We are interested in vertices which have a non-negative mass dimension
and no $v/M_S$-suppression. Thus, the selection of effective vertices we have to inspect is finite.
We further focus on contributions with minimal powers in the Yukawa coupling and maximal powers in the QCD gauge coupling. For such contributions we inspect the 
highest power $x_q$ contributions which are unsuppressed i.e.~$\propto (v/M_S)^0$.

\subsection{Effective vertices with fermions and the quartic interaction of
  Higgs bosons}

The procedure of a LME naturally produces the structure of an EFT. The
effective vertices arise from applying Taylor operations on
one-light-particle irreducible (1LPI) diagrams.
Here we carry out a dimensional analysis of the effective vertices
arising in this way, with focus on the appearance of $x_q$.
Symbolically our notation of the expressions is given as
\begin{align}
\label{eq:taylor_integrand}
\begin{split}
\mathcal{T}_{\left\{m,p\right\}} \left\{
\begin{tikzpicture}[baseline=(c)]
\begin{scope} [shift={(0.,0)}]
\begin{scope} [shift={(-1.5,0)}]
\draw[fermionnoarrow] (1,0) arc (180:0:.5);
\draw[fermionnoarrow] (2,0) arc (0:-180:.5);
\node at (1.5,0){};
\end{scope}
%
%filling
\begin{scope}
\clip (0,0) circle (.5cm);
\foreach \x in {-.9,-.8,...,.6}
\draw[line width=1 pt] (\x,-.6) -- (\x+.6,.6);
\end{scope}
\node at (145:1.65) {$l$};
\node at (215:1.7) {$l$};
\node at (35:1.65) {$l$};
\node at (-35:1.7) {$l$};
\draw[fermionnoarrow](145:1.5) -- (145:.5cm);
\draw[fermionnoarrow](215:1.5) -- (215:.5cm);
\draw[fermionnoarrow](35:1.5) -- (35:.5cm);
\draw[fermionnoarrow](-35:1.5) -- (-35:.5cm);
\node (c) at (0,-0.1){};
\end{scope}	
\end{tikzpicture}
\right\}
{=}&
\begin{tikzpicture}[baseline=(c)]
\begin{scope} [shift={(0.,0)}]
\node at (145:1.65) {$l$};
\node at (215:1.7) {$l$};
\node at (35:1.65) {$l$};
\node at (-35:1.7) {$l$};
\draw[fermionnoarrow](145:1.5) -- (145:.14cm);
\draw[fermionnoarrow](215:1.5) -- (215:.14cm);
\draw[fermionnoarrow](35:1.5) -- (35:.14cm);
\draw[fermionnoarrow](-35:1.5) -- (-35:.14cm);
\draw (-0.1,-0.1) rectangle (0.1,0.1);
\end{scope}
\end{tikzpicture}
\equiv \Delta v_{l^4} \supset x_q^{\leq j}~,
\end{split}
\end{align}
where the blob on the left-hand side denotes a 1LPI Green function
with light external fields $l$ of mass scale $m$ and with external
momenta $p$. The symbol $\mathcal{T}$ denotes the Taylor operator with
respect to the indicated variables,
which by definition acts on the Feynman integrand. On the right-hand side, $\Delta v_{l^4}$ represents the
effective vertex, which is a  power
series in $1/M_S$.
Furthermore, in eq.~\eqref{eq:taylor_integrand} the notation ``$\Delta v \supset x_q^{\leq j}$'' specifies that only terms with $x_q^{\leq j}$ appear  in the effective vertex,
while terms with $x_q^{> j}$ are absent.
 
As outlined in \secref{sec:threshold}, the diagrammatic contributions enhanced by the squark mixing parameter $x_q$ arise in two ways:
\begin{itemize}
	\item by the trilinear coupling of the Higgs bosons  with the left- and right-handed squarks in eq.~\eqref{eq:trilinear_sim} which is accompanied by a factor of the Yukawa coupling
	\item by the chirality flip vertex, induced by 
	the off-diagonal entry in the squark-mass matrix eq.~\eqref{eq:stop_mass_mat}, which is accompanied by a factor of quark mass $m_q\propto m$ and the scale $M_S$.
\end{itemize} 
At leading QCD order there are no internal vertices with a Yukawa
coupling and  
hence no internal Higgs boson propagators. 
Therefore, trilinear squark-Higgs vertices are only possible
at couplings with the external Higgs lines, and the number of possible $x_q$ enhanced trilinear vertices is bounded by 
the number of external Higgs lines associated with the effective
vertex $\Delta v$.
In stark contrast, the chirality flip vertex in the internal squark propagator 
could be inserted arbitrarily often at any fixed loop order. 
Nonetheless, dimensional analysis can be applied in order to establish
a relation between the number of chirality flips and the 
power of the mass suppression factor $m/M_S$ as follows.

For the detailed analysis we fix 
$\Delta v$ to be an effective vertex with $n_h$ external Higgs boson lines of mass dimension $n_d$, i.e.~${\left[\Delta v\right] = \left[\text{mass}\right]^{n_d}}$.
Furthermore, we allow for  $n_{I}$ chirality flip insertions in the squark lines.
For this case, the
highest-power $x_q$ contribution to the effective vertex is given by
\begin{align}
\label{eq:Delta_v_eff}
 \Delta v \supset & \left({y_q} x_q M_S\right)^{n_h} \left( m M_S x_q \right)^{n_{I}}  \int_{k} \mathcal{T}_{\left\{m,p\right\}} f(m,p,M_S,k)~,
\end{align}
where the common appearances of $x_q$ and the Yukawa coupling, the
mass $m$ and $M_S$ has been made explicit. %each Yukawa coupling introduces one factor of $x_q$.
The evaluation of the leading term in the Taylor expansion leads to a loop integral
function which depends  on the physical scale $M_S$ only 
\footnote{We want to stress that terms with less powers in $x_q$ have been neglected.}
\begin{align}
 \label{eq:Delta_v_taylor}
 \Delta v\supset & \left({y_q} x_q M_S\right)^{n_h} \left( m M_S x_q \right)^{n_{I}} p^{n_p} I(M_S)\left[\text{const} + \ord\left( \frac{m,p}{M_S} \right)\right],\\
 \label{eq:Delta_v_mass_supp}
 = & ({y_q} x_q)^{n_h} (x_q m)^{n_{d} -n_p} p^{n_p}
   \left(x_q\frac{m}{M_S}\right)^{n_{I}+n_p-n_{d} }
 \left[\text{const} + \ord\left( \frac{m,p}{M_S}  \right)\right],
\end{align}
where $n_p$ represents the minimum power in the external
momentum $p$ required for the effective vertex (the precise Lorentz
structure of the effective vertex can be ignored and is not specified).
In the last equation we used the fact that $\Delta v$ is of mass dimension $[\text{mass}]^{n_{d}}$.
For ${n_{d}}>n_{I}$ the dimensional analysis does not forbid
 contributions to the effective vertex with positive 
 powers in $M_S$.\footnote{Additional arguments have to be invoked
in order to inspect all diagrammatic contributions to an effective vertex
without spurious $M_S$ enhancement related to chiral symmetry, gauge symmetry or \SUSY.}

The previous equations elucidate that for a fixed vertex function
(with fixed $n_h$, $n_d$, $n_p$), each additional internal squark
chirality flip results in an additional mass suppression factor
$m/M_S$. 
For the especially interesting case $n_p=0$, 
the maximal power of $x_q$ without mass suppression is realized for
${n_{d}}=n_{I}$, and this maximal power is $ x_q^{n_{d}+n_h} \subset \Delta v$.

Eq.~\eqref{eq:Delta_v_eff} implies that a
product of effective
vertices $\Pi_{i=1}^n \Delta v_i$ behaves the same way as the expression on 
the r.h.s. of eq.~\eqref{eq:Delta_v_mass_supp}, i.e.~the product of effective 
vertices can be expressed with indices $n_h'$, $n_I'$ and $n_d'$ obtained 
from the respective sum of the indices $(n_h)_i$, $(n_I)_i$ and $(n_d)_i$ 
of the individual effective vertices $\Delta v_i$. 

We note that the reasoning does not depend on the number of fermions 
coupled to Higgs bosons by the effective vertex $\Delta v$.
Furthermore, the arguments are also independent of the specific order in 
the QCD coupling $g_3$.

In the following we list the results from analyzing
diagrammatic contributions with non-negative mass dimension 
by this procedure. The following results thus correspond to explicit
versions of eq.\ (\ref{eq:Delta_v_mass_supp}), specialized to these vertices.
On the r.h.s.\ we always specify only the leading terms and suppress possible factors 
 $\left[\text{const} + \ord\left( \frac{m,p}{M_S}  \right)\right]$.
As in sec.\ \ref{sec:GF} we will denote effective vertices which contain unsuppressed $x_q$ contributions by gray squares.
In contrast, effective vertices without $x_q$ enhancement are represented by white squares. 

\begin{align}
\label{eq:eff_vert_qqh}
\intertext{(1) effective Higgs-quark vertex $\Delta v_{\bar q qh} $:}
	\begin{tikzpicture}[baseline=(c)]
	\begin{scope} [shift={(0,0)}]	
	\draw[fermion] (0,0)--(1,0);
	\draw[fermion] (1.2,0)--(2.2,0);
	\draw[scalarnoarrow] (1.1,1.1)--(1.1,0.1);
	\begin{scope}[shift={(0,0)}]
	\draw[fill=gray](1.,-0.1) rectangle (1.2,0.1);
	\end{scope}
	\end{scope}
	\node at (.4,.3) {$q$};	
	\node at (1.75,.3) {$q$};
	\node at (1.4,1.) {$h$};
	\end{tikzpicture} 
	&\supset {y_q} g_3^{2k}  x_q^{\leq 1}\left(x_q \frac{m}{M_S}\right)^{n_I} \\
\intertext{(2) effective quark propagator insertion $\Delta v_{\bar q q} $:}
	\begin{split}
	\begin{tikzpicture}[baseline=(c)]
	\begin{scope} [shift={(0.,0)}]
	\draw[fermion] (0,0)--(1,0);
	\draw[fermion] (1.2,0)--(2.2,0);
	\node (c) at (0,-0.1){};
	\begin{scope}[shift={(0,0)}]
	\draw[fill=gray](1.,-0.1) rectangle (1.2,0.1);
	\end{scope}
	\end{scope}
	\node at (.4,.3) {$q$};	
	\node at (1.75,.3) {$q$};
	\end{tikzpicture} 
	&\equiv\Delta v_{\bar qq} = \Delta v_m + \Delta v_{\slashed p}
	\end{split}\\
	\label{eq:eff_vert_qq}
	&\quad~
	\Delta v_m \supset  g_3^{2k}\, m\, x_q^{\leq 1}\left(x_q \frac{m}{M_S}\right)^{n_I-1} \\
	&\quad~
	\Delta v_{\slashed p} \supset  g_3^{2k}  \,\slashed p \,x_q^0\left(x_q \frac{m}{M_S}\right)^{n_I} 
\\
\label{eq:eff_vert_hhhh}
\intertext{(3) effective quartic Higgs vertex $\Delta v_{h^4}$:}
\label{eq:LME_h^4}
	\begin{tikzpicture}[baseline=(c)]
	\begin{scope} [shift={(0.,0)}]
\draw[scalarnoarrow] (135:.1275cm)--(135:1.08)  ;
\node at (145:1.) {$h$};
\draw[scalarnoarrow](225:.1275cm)--(225:1.08) ;
\node at (215:1.) {$h$};
\draw[scalarnoarrow] (45:.1275cm) -- (45:1.08);
\node at (35:1.) {$h$};
\draw[scalarnoarrow] (-45:.1275cm) --(-45:1.08);
\node at (-35:1.) {$h$};
	\begin{scope}[shift={(-1.1,0)}]
	\draw[fill=gray](1.,-0.1) rectangle (1.2,0.1);
	\end{scope}
	\end{scope}
	\end{tikzpicture} 
	&\supset  y_q^4 g_3^{2k}  x_q^{\leq 4}\left(x_q \frac{m}{M_S}\right)^{n_I} \\
\intertext{(4) effective gluon-quark vertex $\Delta v_{g\bar qq} $:}
	\begin{split}
	\begin{tikzpicture}[baseline=(c)]
	\begin{scope} [shift={(0.,0)}]
	\draw[fermion] (0,0)--(1,0);
	\draw[fermion] (1.2,0)--(2.2,0);
	\draw[gluon] (1.1,0.1)--(1.1,1.05);
	\draw (1.,-0.1) rectangle (1.2,0.1);
	\end{scope}
	\node at (.4,.3) {$q$};	
	\node at (1.75,.3) {$q$};
	\node at (1.4,1.) {$g$};
	\end{tikzpicture}
	&\supset g_3^{2k+1}  x_q^0 \left[ \gamma^{\mu} + \ord\left( \frac{p^\mu, m \gamma^\mu}{M_S}  \right) \right]\left(x_q \frac{m}{M_S}\right)^{n_I} .
	\end{split}\\
\intertext{(5) effective vertex for gluon interactions $\Delta v_{g^3}$ and $\Delta v_{g^4}$: By covariant decomposition
	and dimensional analysis it follows that $x_q$ enhanced contributions are suppressed as}
	\begin{tikzpicture}[baseline=(c)]
	\node at (.4,.3) {$g$};	
	\node at (1.75,.3) {$g$};
	\node at (1.4,1.) {$g$};
	\begin{scope} [shift={(0.,0)}]
	\draw[gluon] (0.05,0)--(1,0);
	\draw[gluon] (1.2,0)--(2.15,0);
	\draw[gluon] (1.1,.1)--(1.1,1.05);
	\draw (1.,-0.1) rectangle (1.2,0.1);
	\end{scope}
	\end{tikzpicture}& \supset g_3^{2k+1}  \left(\eta^{\mu\nu} p^\sigma\right) 
	\left(x_q \frac{m}{M_S}\right)^{n_I},
	\\
	\begin{tikzpicture}[baseline=(c)]
	\begin{scope} [shift={(0.,0)}]
	\draw[gluon] (135:.135cm)--(135:1.09)  ;
	\node at (152:1.) {$g$};
	\draw[gluon](225:.135cm)--(225:1.09) ;
	\node at (213:1.) {$g$};
	\draw[gluon] (45:.135cm) -- (45:1.09);
	\node at (30:1) {$g$};
	\draw[gluon] (-45:.135cm) --(-45:1.09);
	\node at (-25:1.) {$g$};
		\begin{scope}[shift={(-1.1,0)}]
	\draw(1.,-0.1) rectangle (1.2,0.1);
	\end{scope}
	\end{scope}
	\end{tikzpicture} 
	&\supset  g_3^{2k+1}  \eta^{\mu \nu} \eta^{\sigma \rho} 
	\left(x_q \frac{m}{M_S}\right)^{n_I}.
\end{align}

\subsection{Effective vertices with gluons, ghosts and Higgs bosons}

For the following effective vertices involving external gluons and Higgs bosons, we extend the reasoning by arguments based on BRST invariance and its consequences.
To exemplify why this is helpful, we consider a diagrammatic
contribution to the gluon propagator $\Gamma_{g g}$.  
Based on a Taylor expansion (at leading order) and dimensional analysis one might evaluate individual diagrams as

\begin{align}
\label{eq:vgg_MS}
\begin{tikzpicture}[baseline=(c)]
\begin{scope} [shift={(0.,0)}]	
\draw[gluon] (0.05,0)--(1,0);
\draw[fermionnoarrow] (1,0) arc (180:0:.5);
\draw[fermionnoarrow] (2,0) arc (0:-180:.5);
\draw[gluon] (2,0)--(2.95,0);
\node at (1.5,0){\scriptsize{$\mathcal{T}$}};
\node (c) at (0,-0.1){};
\node at (.5,.3) {$g$};	
\node at (2.5,.3) {$g$};
\end{scope}
\end{tikzpicture}
&  
\begin{tikzpicture}[baseline=(c)] 
\end{tikzpicture} \supset g_3^{2k} (M_S+m)^2 \eta^{\mu \nu} +\ord(p^\mu p^\nu, \eta^{\mu\nu} p^2),
\end{align}
where the $\cal  T$ denotes that the corresponding loop integrand is Taylor expanded in $m$ and $p$.
The leading term in the LME absorbs  UV contributions from heavy fields ($M_S \gg m$)
in the effective vertices of light fields present in the EFT.
Thus, the effective vertex can be regarded as a 
construction of decoupling coefficients
of operators in the EFT Lagrangian of the SM which in turn is invariant under $SU(3)_{C}$
BRST transformations.

Individual diagrams indeed behave in the way shown in eq.~\eqref{eq:vgg_MS}. However, after the inclusion of all diagrams to $\Gamma_{g g}$ the entire contribution has to be proportional to the transverse projector $T^{\mu \nu}\propto p^\mu p^\nu -p^2 \eta^{\mu\nu}$.
Relying on the existence of a BRST invariant EFT implies
transversality also after Taylor expansion, i.e.\ 
the full result in eq.~\eqref{eq:vgg_MS} is independent of the $(M_S
+m)^2$ term  at each order,
\begin{align}
\label{eq:gluon-2-vertex}
	\sum\limits_{\text{heavy lines}\geq 1}
\begin{tikzpicture}[baseline=(c)]
\begin{scope} [shift={(0.,0)}]	
\draw[gluon] (0.05,0)--(1,0);
\draw[fermionnoarrow] (1,0) arc (180:0:.5);
\draw[fermionnoarrow] (2,0) arc (0:-180:.5);
\draw[gluon] (2,0)--(2.95,0);
\node at (.5,.3) {$g$};	
\node at (2.5,.3) {$g$};
\node at (1.5,0){\scriptsize{$\mathcal{T}$}};
\node (c) at (0,-0.1){};
\end{scope}
\end{tikzpicture}
&\overset{\phantom{\text{LME}}}{\equiv} 
\begin{tikzpicture}[baseline=(c)]
\begin{scope} 	
\draw[gluon] (0.05,0)--(1,0);
\draw[gluon] (1.2,0)--(2.15,0);
\node at (.4,.3) {$g$};	
\node at (1.75,.3) {$g$};
\draw (1.,-0.1) rectangle (1.2,0.1);
\end{scope}
\end{tikzpicture} 
	\propto g_3^{2k} T^{\mu \nu}.
\end{align}

In the following we continue our list of effective
vertices. In each case we now assume a  summation over all
contributing diagrams to ensure the cancellation of spurious
contributions analogous to the ones on the r.h.s.~in eq.~\eqref{eq:vgg_MS}.
\begin{align}
	\intertext{(6) gluon-gluon effective vertex $\Delta v_{gg}$  : 
	Having discussed the transverse structure of the effective vertex, we can
	link the contributions in eq.~\eqref{eq:gluon-2-vertex} to the master formula
	in eq.~\eqref{eq:Delta_v_mass_supp}. In that formula the
        transversality now implies the momentum dependence
        $n_p\geq2$, and as a result the  $x_q$ appearance can be
        characterized as}
	\begin{tikzpicture}[baseline=(c)]
	\begin{scope} 	
	\draw[gluon] (0.05,0)--(1,0);
	\draw[gluon] (1.2,0)--(2.15,0);
	\draw (1.,-0.1) rectangle (1.2,0.1);
	\end{scope}
	\node at (.4,.3) {$g$};	
	\node at (1.75,.3) {$g$};
	\end{tikzpicture}
	&
	\supset g_3^{2k}  T^{\mu \nu} \left(x_q \frac{m}{M_S}\right)^{n_I}.
\intertext{(7) effective Higgs-gluon vertex $\Delta v_{ggh}$  and $\Delta v_{gghh}$ :
The Higgs-gluon interactions can be associated to 
dimension 6 (or higher) operators in the corresponding EFT and as
before  we obtain $n_p\geq2$ 
	in eq.~\eqref{eq:Delta_v_mass_supp}. Therefore, these
        effective interactions are constrained as follows,}
	\begin{tikzpicture}[baseline=(c)]
\node at (.4,.3) {$g$};	
\node at (1.75,.3) {$g$};
\node at (1.3,1.) {$h$};
\begin{scope} [shift={(0.,0)}]
\draw[gluon] (.05,0)--(1,0);
\draw[gluon] (1.2,0)--(2.15,0);
\draw[scalarnoarrow] (1.1,1.1)--(1.1,0.1);
\draw (1.,-0.1) rectangle (1.2,0.1);
\end{scope}
\end{tikzpicture}
	&\supset {y_q} g_3^{2k}  (p^2)^{\mu \nu}  \left(\frac{1}{M_S}\right)
	\left(\frac{x_q m_q}{M_S}\right)^{n_I} ,\\
	\begin{tikzpicture}[baseline=(c)]
\begin{scope} [shift={(0.,0)}]
\draw[gluon] (135:1.09) --(135:.135cm)  ;
\node at (148:1.) {$g$};
\draw[gluon] (225:1.09)--(225:.135cm) ;
\node at (210:1.) {$g$};
\draw[scalarnoarrow] (45:.135cm) -- (45:1.09);
\node at (30:1) {$h$};
\draw[scalarnoarrow] (-45:.135cm) --(-45:1.09);
\node at (-25:1.) {$h$};
\begin{scope}[shift={(-1.1,0)}]
\draw(1.,-0.1) rectangle (1.2,0.1);
\end{scope}
\end{scope}
\end{tikzpicture} 
	&
	\supset  y_q^2 g_3^{2k}  (p^2)^{\mu \nu}   \left(\frac{1}{M_S^2}\right)
	\left(\frac{x_q m_q}{M_S}\right)^{n_I},\\
\intertext{where $(p^2)^{\mu \nu} $ represents a vertex-dependent bilinear combination of
	the external momenta. Importantly, both contributions are
        suppressed by large $M_S$.}
	\intertext{(8) effective ghost-antighost vertex $\Delta v_{\bar cc}$:
	The effective vertices involving ghosts can be discussed in a similar way.
	Due to BRST invariance, the effective QCD ghost-antighost vertex $\Delta v_{\bar cc}$  obeys a proportionality to the squared external
	momentum $p^2$, resulting in suppressed $x_q$-enhanced contributions}
	\begin{tikzpicture}[baseline=(c)]
\begin{scope} 	
\node at (.4,.3) {$c$};	
\node at (1.65,.3) {$c$};
\draw[ghost] (0,0)--(1,0);
\draw[ghost] (1.2,0)--(2.2,0);
\draw (1.,-0.1) rectangle (1.2,0.1);
\end{scope}
\end{tikzpicture} &
		\supset g_3^{2k} p^2\left(\frac{x_q m}{M_S}\right)^{n_I}.
\intertext{(9)
	effective ghost-gluon vertex $\Delta v_{g\bar cc}$: This effective vertex can be directly evaluated by
	covariant decomposition and dimensional analysis as}
	\begin{tikzpicture}[baseline=(c)]
	\begin{scope} 	
	\draw[ghost] (0,0)--(1,0);
	\draw[ghost] (1.2,0)--(2.2,0);
	\draw[gluon] (1.1,1.05)--(1.1,0.1);
	\draw (1.,-0.1) rectangle (1.2,0.1);
	\node at (.4,.3) {$c$};	
	\node at (1.65,.3) {$c$};
	\node at (1.4,1.) {$g$};
	\end{scope}
	\end{tikzpicture}& \supset g_3^{2k +1} p^\mu\left(\frac{x_q m}{M_S}\right)^{n_I}.
\end{align}
There are further QCD related interactions, trilinear and quartic, involving Higgs, ghosts and gluons, which we do not list.
Because of BRST invariance such interactions are associated with higher dimensional operators in the EFT and are necessarily suppressed at least by a factor of $p/M_S$. 

In summary, using eq.~\eqref{eq:Delta_v_mass_supp} we have established the properties that are necessary for classification of
 eqs.~(\ref{eq:higher_dim},\ref{eq:effvert_lowdim},\ref{eq:list}):
 \begin{itemize}
  \item The crucial behavior of the effective vertices in eq.~\eqref{eq:higher_dim}
 	is characterized by ${n_d<0}$. 
 	Eq.~\eqref{eq:Delta_v_mass_supp} implies, irrespectively
 	of the specific $x_q$ appearance, that
 	any contribution is suppressed at least by a factor of $v/M_S$. 
 	\item The effective vertices in eq.~\eqref{eq:effvert_lowdim}
 	are described by $n_d=n_p$ and $n_h=0$. Therefore, no positive $x_q$ power can originate at $\ord((v/M_S)^0)$. 
 	\item The  central effective vertices in eq.~\eqref{eq:list} which 
 	introduce positive powers of $x_q$ arise in two ways.
 	On the one hand the two-quark vertex in eq.~\eqref{eq:eff_vert_qq} has the property $n_d>n_p$
 	and on the other hand the effective vertices in eqs.~(\ref{eq:eff_vert_qqh},\ref{eq:eff_vert_hhhh}) feature $n_h\geq0$ and  $n_d=n_p$.
\end{itemize}
\newpage
\addcontentsline{toc}{section}{References}

\bibliographystyle{JHEP}
\bibliography{xf_theorem}

\end{document}